\documentclass[prd,aps,nofootinbib,notitlepage]{revtex4}
\usepackage{graphicx,epsf,amsmath,amsfonts,amssymb,amsbsy}
\usepackage{epsfig}
\usepackage{epstopdf}
\newcommand{\ds}{\displaystyle}
\newcommand{\vev}[1]{\langle#1\rangle}
\newcommand{\mat}{\left ( \begin{array}}
\newcommand{\emat}{\end{array} \right )}
\newcommand{\vect}{\left ( \begin{array}{c}}
\newcommand{\evect}{\end{array} \right )}

\newcommand{\Det}{\mathop{\rm Det}\nolimits}
\begin{document}

\title{ \bf 
Charged pion condensation and duality in dense and hot chirally and isospin asymmetric quark matter in the framework of NJL$_2$ model}

\author{T. G. Khunjua $^{1)}$, K. G. Klimenko $^{2)}$, and R. N. Zhokhov $^{2),~3)}$ } 
\vspace{1cm}

\affiliation{$^{1)}$ Faculty of Physics, Moscow State University,
119991, Moscow, Russia}
\affiliation{$^{2)}$ State Research Center
of Russian Federation -- Institute for High Energy Physics,
NRC "Kurchatov Institute", 142281, Protvino, Moscow Region, Russia}
\affiliation{$^{3)}$  Pushkov Institute of Terrestrial Magnetism, Ionosphere and Radiowave Propagation (IZMIRAN),
108840 Troitsk, Moscow, Russia}


\begin{abstract}
In this paper we investigate in the large-$N_c$ limit ($N_c$ is the number of colored quarks) the phase structure of a
massless (1+1)-dimensional quark model with four-quark interaction
and in the presence of baryon ($\mu_B$), isospin ($\mu_I$) and chiral isospin ($\mu_{I5}$) chemical potentials 
as well as at nonzero temperature. It is established that chiral isospin  chemical potential leads to the generation of 
charged pion condensation (PC) in dense 
(nonzero baryon density) and chiral asymmetric quark matter for a wide range of isospin densities. It is shown that 
there exists a duality correspondence
between the chiral symmetry breaking and the charged PC phenomena at any values of temperature even for very hot quark gluon plasma. 
Moreover, it is shown that charged PC phase with nonzero baryon density can be induced in the model at comparatively high 
temperatures. This opens up new possible physical systems, where it can be of importance, such as heavy ion collisions, just born 
neutron stars (proto-neutron stars), supernovas as well as neutron star mergers. 

\end{abstract}

\maketitle

\section{Introduction}

Recently, much attention has been paid to the study of dense baryon (quark) medium with isotopic (isospin) asymmetry 
(different densities of $u$ and $d$ quarks). Such matter can exist inside compact stars, it can appear in heavy-ion collision experiments, etc, and it is usually described in terms of different nonperturbative QCD methods or effective QCD-like theories such as chiral
effective Lagrangians or, especially,  Nambu -- Jona-Lasinio (NJL)
type models \cite{njl} with nonzero baryon $\mu_B$ and isospin $\mu_I$ chemical potentials. It turns out that the 
$(\mu_I,\mu_B)$ phase diagram of these models gives us the opportunity to better understand phenomena such as chiral symmetry 
restoration \cite{asakawa,ebert,sadooghi,hiller,boer}, color superconductivity \cite{alford,klim,incera}, and charged pion 
condensation (PC) effect \cite{son, he,eklim,ak,mu,andersen,ekkz,gkkz,thiesmu5, kkz1+1,Mammarella:2015pxa,Carignano:2016lxe,symmetry} (let us also recall ideas about pion stars \cite{Andersen:2018nzq})   that can be observed in
dense quark matter with isotopic asymmetry.

While the reality of the first two of the above phenomena is beyond doubt, the possibility of condensation of charged pions
in dense quark medium is not reliably established (at least in the framework of the NJL model considerations). Indeed, for
some values of the model parameters (the coupling constant $G$, the cutoff parameter $\Lambda$, etc.), the phase of quark 
matter with {\it nonzero baryon density}, in which the charged pions are condensed, is allowed by the NJL models. However, 
this effect is prohibited by the NJL models for other physically interesting $G$ and $\Lambda$ values \cite{eklim}. In 
addition, if the condition of electric neutrality is superimposed on quark matter, then the charged PC phenomenon strongly 
depends on the size of the bare quark mass $m_0$. In particular, it turns out that in the framework of the NJL models 
the condensation of charged pions is forbidden in medium with {\it nonzero baryon density} if $m_0$ reaches physically 
acceptable values of $5\div 10$ MeV (see in Ref. \cite{andersen}). However, on the basis of the toy (1+1)-dimensional model
with baryon and isospin chemical potentials, it was shown in Refs. \cite{ekkz,gkkz} that there are factors (the spatial 
sizes of the system or the spatial inhomogeneity of its condensates) that can stimulate the appearance of the charged PC 
phenomenon in dense quark matter.

Quite recently, it was found that in dense quark medium in the presence of an external magnetic field two other effects, chiral 
magnetic and chiral separation effects \cite{fukus,Metlitski}, can be observed. So they can be realized in compact stars 
and  heavy-ion collision experiments, etc. Usually, these phenomena are inherent in such environments in which there is a 
chiral imbalance (or asymmetry), i.e. when in dense quark matter there is a difference between the density $n_R$ of all
right-handed and the density $n_L$ of all left-handed quarks. The quantity $n_{5}\equiv n_R-n_L$ is called the chiral 
density of the system. It has been argued that the chiral density $n_5$ can be
generated dynamically at high temperatures, for example, in the fireball after heavy ion collision, by virtue of the 
Adler-Bell-Jackiw anomaly and quarks interacting with gauge (gluon) field configurations with nontrivial topology, named sphalerons.  
In the presence of an external strong magnetic field, which can be produced in heavy ion collisions as well, this can lead to 
the so-called chiral magnetic effect \cite{fukus}. Moreover, in the presence of external magnetic field 
chiral density 
$n_5$ can be produced (even at rather low temperature) in dense quark matter due to the so-called chiral separation  
effect \cite{Metlitski} (it can be also produced under fast rotations of the system due to the so-called chiral vortical effect).  
Now let us notice that usually when one talk about chiral density $n_5$ one implies that chiral density $n_{u5}$ of 
$u$ quarks and chiral density $n_{d5}$ of $d$ quarks are equal to each other (it is evident that $n_{5}=n_{5u}+n_{5d}$). Indeed, that 
is the case when one has in mind the mechanism of generation of 
chiral imbalance at high temperatures due to nontrivial topology of gauge field configuration. In this case it is quite plausible 
that $n_{u5}=n_{d5}$ due to the fact that gluon field interacts with different quark flavours in exactly the same way and does not feel 
the difference between flavours.
But another mechanism, the chiral separation effect, is sensitive to the flavour of quarks (as it was shown in Appendices A to 
the papers \cite{kkz2,kkzjhep}). So in dense quark matter ($\mu_B\ne 0$) a strong magnetic field separates $u$ and $d$ quarks in different ways. As a 
result, we see that, e.g., in such astrophysical objects as magnetars there might exist areas, in which the quantity $n_{I5}=n_{5u}-
n_{5d}$, called the chiral isospin density, is not zero. Moreover, it has been argued that chiral imbalance is generated by parallel magnetic and electric fields \cite{Ruggieri:2016fny}, one can generalize these arguments to chiral isospin imbalance as well. 

So in the most general case, chiral imbalance is described by two chemical potentials, 
chiral $\mu_5$ and chiral isospin (or isotopic) $\mu_{I5}$ chemical potentials, which are thermodynamically conjugated to 
$n_{5}$ and $n_{I5}$, respectively. The first, $\mu_5$, is usually used when isotopic 
asymmetry of quark matter is absent, i.e. in the case $\mu_I=0$ \cite{andrianov,braguta}. The second, $\mu_{I5}$, might be taken 
into account when, in addition to chiral, there is also isotopic asymmetry of matter, in which charged PC phenomenon may occur, etc. 
\cite{kkz}. In particular, it was established in the framework of NJL models that $\mu_5$ catalyzes the chiral symmetry 
breaking (CSB) \cite{braguta,kkz}, whereas $\mu_{I5}$ promotes the charged PC in dense quark matter \cite{kkz}.  

It was also shown 
in Refs. \cite{kkz} that in the leading order of the $1/N_c$ expansion the most general $(\mu_B,\mu_I,\mu_5,\mu_{I5})$-phase 
portrait of the massless NJL model is symmetric with respect to a so-called dual transformation between charged PC 
and CSB phases. If the bare quark mass $m_0$ is not zero, then this duality is only approximate, although quite accurate, 
symmetry of the NJL model phase portrait (see in Ref. \cite{kkz2}). In addition, it was established in this paper that  
chiral isospin chemical potential $\mu_{I5}$ promotes charged PC phenomenon in dense quark matter even at physically 
acceptable values of $m_0$.

One of the drawbacks of the (3+1)-dimensional NJL (NJL$_4$) model is its non-renormalizability. Therefore, the predictions of this 
effective model depend on the cutoff parameter, which is typically chosen to be of the order of 1 GeV. So, the results of 
the NJL$_4$ model usage are valid only at {\it comparatively low energies, temperatures and densities} (chemical potentials).
However, there exists also a class of renormalizable theories, the (1+1)-dimensional chiral Gross--Neveu (GN) type models 
\cite{gn,ft}, \footnote{Below we shall use the notation ``NJL$_2$ model''  instead of ``chiral GN model'' for 
(1+1)-dimensional models with {\it continuous chiral and/or isotopic, etc, symmetries}, since the chiral structure of their
Lagrangians is the same as that of the corresponding (3+1)-dimensional NJL models.} that can be used as a laboratory for 
the qualitative simulation of specific properties of QCD at {\it arbitrary energies}. Renormalizability, asymptotic 
freedom, as well as the spontaneous chiral symmetry breaking (in vacuum) are the most fundamental inherent features both 
for QCD and all NJL$_2$ type models. In addition, the $\mu_B-T$ phase diagram (where $T$ denotes temperature) is 
qualitatively the same both for the QCD and NJL$_2$ models \cite{wolff,kgk1,barducci,chodos}. Let us further mention that
(1+1)-dimensional Gross-Neveu type models are also suitable for the description of physics in quasi one-dimensional 
condensed matter systems like polyacetylene \cite{caldas}. It is currently well understood (see, e.g., the discussion in Refs. \cite{barducci,chodos,thies}) that the usual {\it no-go} theorem \cite{coleman}, which generally forbids the spontaneous breaking of any continuous symmetry in two-dimensional spacetime, does not work in the limit  $N_c\to\infty$, where $N_c$ is the number of colored quarks. This follows directly from the fact that in the limit of large $N_c$ the quantum fluctuations, which would otherwise destroy a long-range order corresponding to a spontaneous symmetry breaking, are suppressed by $1/N_c$ factors. Thus,  the effects
inherent for real dense quark matter, such as CSB phenomenon  (spontaneous breaking of the continuous axial $U(1)$
symmetry) or charged pion condensation (spontaneous breaking of the continuous isospin symmetry) might be simulated in terms of a 
simpler NJL$_2$-type models, though only in the leading order of
the large-$N_c$ approximation (see, e.g., Refs. \cite{thies} and \cite{ektz,massive,ek2,adhikari,gubina}, respectively).

This paper is devoted to the investigation of such phenomena of dense quark matter as CSB and charged PC, as well as their 
mutual influence on each other, in the framework of an extended toy NJL$_2$ model with two massless quark flavors and 
in the presence of the baryon $\mu_B$, isospin $\mu_I$ as well as chiral isospin $\mu_{I5}$ chemical potentials 
(for simplicity, we consider only the case $\mu_5=0$). Of course, in this case, in two dimensions, a significant part of 
physical processes that can occur in real dense quark medium in (3 + 1) dimensions, falls out of our consideration. 
However, as discussed, e.g., in Ref. \cite{Gusynin:1994xp}, the dynamics of the pairing of fermions in strong magnetic 
fields is (1 + 1) dimensional. Therefore, such phenomena of real dense quark matter as spontaneous breaking of chiral symmetry, 
condensation of charged pions and other pairing phenomena in strong magnetic fields can be investigated effectively in 
the framework of two-dimensional models. The chemical potential $\mu_{I5}$ is also included in our consideration, 
because the regions with $\mu_{I5}\ne 0$ can appear (due to the chiral separation effect) in neutron stars or arise in 
quark matter formed by the collision of heavy ions just under the influence of strong magnetic field 
(this fact is discussed in Appendix A of Refs. \cite{kkz2,kkzjhep}).
Moreover, in order to avoid the {\it no-go} theorem \cite{coleman}, we perform all calculations in the leading order of 
the large-$N_c$ technique. To clarify the true role of the chiral isospin $\mu_{I5}$ chemical potential in the creation of 
the CSB and charged PC in dense quark matter, it is supposed throughout the paper that all condensates are spatially 
homogeneous. \footnote{As it was noted above, spatial inhomogeneity of condensates by itself can cause charged PC in dense 
baryon matter \cite{gkkz}.} Under this constraint the model was already investigated in Refs. \cite{ektz,massive} 
at $\mu_{I5}=0$, where it was shown that the charged PC phase with nonzero baryon density is forbidden at arbitrary values 
of $\mu_B$ and $\mu_I$ (at zero chiral imbalance). In contrast, we show that at $\mu_{I5}\ne 0$, i.e. when there is an 
isotopic chiral imbalance of the system, the charged PC phase with nonzero baryon density is allowed to exist for a wide 
range of isospin densities.  

In addition, we show that in the leading order of the large-$N_c$ approximation there arises a duality between CSB and 
charged PC phenomena in the framework of the NJL$_2$ model under consideration for the case of as cold quark matter as well
as for hot media. It means that if at $\mu_I=A$ and $\mu_{I5}=B$ (at arbitrary fixed chemical potential $\mu_B$), e.g., 
the CSB (or the charged PC) phase is realized in the model, then at the permuted values of  these chemical potentials, 
i.e. at $\mu_I=B$ and $\mu_{I5}=A$, the charged PC (or the CSB) phase is arranged. So, it is enough to know the phase 
structure of the model at $\mu_I<\mu_{I5}$, in order to establish the phase structure at $\mu_I>\mu_{I5}$ and vice versa. 
Knowing condensates and other dynamical and thermodynamical quantities of the system, e.g. in the CSB phase, one can then obtain 
the corresponding quantities in the dually conjugated charged PC  phase of the model, by simply performing there the 
duality transformation, $\mu_I\leftrightarrow\mu_{I5}$. \footnote{Note that another kind of duality correspondence, the 
duality between CSB and superconductivity, was demonstrated both in (1+1)- and (2+1)-dimensional NJL models 
\cite{thies2,ekkz2}. } Earlier, the similar dualities between charged PC and chiral symmetry breaking phenomena has been 
observed in the framework of orbifold equivalence formalism in the limit of large $N_c$ \cite{Hanada:2011jb}. Namely, it 
was shown that the whole phase diagram of QCD at chiral chemical potential must be identical to that of QCD at isotopic 
chemical potential in the chiral limit, where the charged pion condensation is replaced by the chiral condensation.

Note that a phase structure of the NJL$_2$ model under consideration with nonzero values of $\mu_{B}$, $\mu_{I}$, and 
$\mu_{I5}$ has been already studied in Ref. \cite{ekk} at $T=0$, but an elusive error has been made there. 
Formally, in Ref. \cite{ekk} all calculations, both numerical and analytical, are correct but one of the statements in 
Appendix, which only at first glance may seem to be correct, turned out to be untrue, and this led to an incorrect  
expression for the thermodynamic potential of the model (see below). However, interestingly enough that the basic 
qualitative conclusions stay true, but some important details of the phase portrait have changed. The predictions 
became stronger and more interesting for the application to real physical scenarios. This paper is devoted to the 
investigation and the analysis of these changes, besides we consider the influence of the temperature on the phase 
diagram and the duality of the model in this case.

The paper is organized as follows. In Sec. II a toy (1+1)-dimensional massless NJL-type model
with two quark flavors ($u$ and $d$ quarks) and with three kinds
of chemical potentials, $\mu_B,\mu_I,\mu_{I5}$, is presented. Next,
we discuss the symmetries of the model under consideration and unrenormalized expression for the 
thermodynamic potential (TDP) both for the case of zero and nonzero temperature is
obtained in the leading order of the large-$N_c$ expansion. Here the dual symmetry of the model TDP
is established at $T\ge 0$. Sec. III is devoted to the calculation of the TDP at $T=0$ both at zero and nonzero values 
of the chemical potentials.
Then, in Sec. IV, the phase structure of the model is investigated at $T=0$. In this section, on the basis of several 
phase diagrams, we demonstrate the existence of a duality between CSB and charged PC phenomena as well the fact that 
$\mu_{I5}$ promotes the realization of the charged PC phase in dense quark matter. It means that charged PC phase 
with nonzero baryon density can be realized in this model only at $\mu_{I5}>0$. Similar results are obtained 
in Sec. V, where the case of nonzero temperature is considered. In the last Sec. VI main results and conclusions of 
the paper are formulated, at the same time, we emphasize the similarity of the phase properties of the NJL model 
in two and four spacetime dimensions with isotopic chiral asymmetry. Some technical
details are relegated to Appendices A and B.

\section{ The model and its thermodynamic potential}

\subsection{The zero temperature case}

We consider in the chiral limit a (1+1)-dimensional NJL model in order to mimic the phase structure of real dense quark matter with two massless quark flavors ($u$ and $d$ quarks). Its Lagrangian, which is symmetrical under global color SU($N_c$) group, has the form
\begin{eqnarray}
&&  L=\bar q\Big [\gamma^\nu\mathrm{i}\partial_\nu
+\frac{\mu_B}{3}\gamma^0+\frac{\mu_I}2 \tau_3\gamma^0+\frac{\mu_{I5}}2 \tau_3\gamma^0\gamma^5\Big ]q+ \frac
{G}{N_c}\Big [(\bar qq)^2+(\bar q\mathrm{i}\gamma^5\vec\tau q)^2 \Big
],  \label{1}
\end{eqnarray}
where the quark field $q(x)\equiv q_{i\alpha}(x)$ is a flavor doublet
($i=1,2$ or $i=u,d$) and color $N_c$-plet ($\alpha=1,...,N_c$) as
well as a two-component Dirac spinor (the summation in (\ref{1})
over flavor, color, and spinor indices is implied); $\tau_k$
($k=1,2,3$) are Pauli matrices in two-dimensional flavor space. The Dirac $\gamma^\nu$-matrices ($\nu=0,1$) and $\gamma^5$ in (1) are matrices in 
two-dimensional spinor space,
\begin{equation}
\begin{split}
\gamma^0=
\begin{pmatrix}
0&1\\
1&0\\
\end{pmatrix};\qquad
\gamma^1=
\begin{pmatrix}
0&-1\\
1&0\\
\end{pmatrix};\qquad
\gamma^5=\gamma^0\gamma^1=
\begin{pmatrix}
1&0\\
0&{-1}\\
\end{pmatrix}.
\end{split}
\end{equation}
Note that at $\mu_{I5}=0$ the model was already investigated in details, e.g., in Refs \cite{ektz,massive,ek2,gubina}. It is evident that the model (\ref{1}) is a generalization of the two-dimensional GN model \cite{gn} with a single massless quark color $N_c$-plet to the case of two quark flavors
and additional baryon $\mu_B$, isospin $\mu_I$ and axial isospin $\mu_{I5}$ chemical potentials. These parameters are introduced in order to describe in the framework of the model (1) quark matter with nonzero baryon $n_B$, isospin $n_I$ and axial isospin $n_{I5}$ densities, respectively.
It is evident that Lagrangian (1), both at $\mu_{I5}=0$ and $\mu_{I5}\ne 0$, is invariant with respect to the abelian $U_B(1)$, $U_{I_3}(1)$ and $U_{AI_3}(1)$ groups, where \footnote{\label{f1,1}
Recall for the following that~~
$\exp (\mathrm{i}\beta\tau_3/2)=\cos (\beta/2)
+\mathrm{i}\tau_3\sin (\beta/2)$,~~~~
$\exp (\mathrm{i}\omega\gamma^5\tau_3/2)=\cos (\omega/2)
+\mathrm{i}\gamma^5\tau_3\sin (\omega/2)$.}
\begin{eqnarray}
U_B(1):~q\to\exp (\mathrm{i}\alpha/3) q;~
U_{I_3}(1):~q\to\exp (\mathrm{i}\beta\tau_3/2) q;~
U_{AI_3}(1):~q\to\exp (\mathrm{i}
\omega\gamma^5\tau_3/2) q.
\label{2001}
\end{eqnarray}
(In (\ref{2001}) the real parameters $\alpha,\beta,\omega$ specify an
arbitrary element of the $U_B(1)$, $U_{I_3}(1)$ and $U_{AI_3}(1)$ groups, respectively.) So the quark bilinears $\frac 13\bar q\gamma^0q$, $\frac 12\bar q\gamma^0\tau^3 q$ and $\frac 12\bar q\gamma^0\gamma^5\tau^3 q$ are the zero components of corresponding conserved currents. Their ground state expectation values are just the baryon $n_B$, isospin $n_I$ and chiral (axial) isospin $n_{I5}$ densities of quark matter, i.e. $n_B=\frac 13\vev{\bar q\gamma^0q}$, $n_I=\frac 12\vev{\bar q\gamma^0\tau^3 q}$
and $n_{I5}=\frac 12\vev{\bar q\gamma^0\gamma^5\tau^3 q}$. As usual, the quantities $n_B$, $n_I$ and $n_{I5}$ can be also found by differentiating the thermodynamic potential of the system with respect to the corresponding chemical potentials. The goal of the
present paper is the investigation of the ground state properties and phase structure of the system (1) and its dependence on the chemical potentials  $\mu_B$, $\mu_I$ and $\mu_{I5}$.

To find the thermodynamic potential of the system, we use a semi-bosonized version of the Lagrangian (\ref{1}), which contains composite bosonic fields $\sigma (x)$ and $\pi_a (x)$ $(a=1,2,3)$ (in what follows, we use the notations $\mu\equiv\mu_B/3$, $\nu\equiv\mu_I/2$ and $\nu_{5}\equiv\mu_{I5}/2$):
\begin{eqnarray}
\widetilde L\ds &=&\bar q\Big [\gamma^\rho\mathrm{i}\partial_\rho
+\mu\gamma^0 + \nu\tau_3\gamma^0+\nu_{5}\tau_3\gamma^0\gamma^5-\sigma
-\mathrm{i}\gamma^5\pi_a\tau_a\Big ]q
 -\frac{N_c}{4G}\Big [\sigma\sigma+\pi_a\pi_a\Big ].
\label{2}
\end{eqnarray}
In (\ref{2}) the summation over repeated indices is implied.
From the Lagrangian (\ref{2}) one gets the Euler--Lagrange equations
for the bosonic fields
\begin{eqnarray}
\sigma(x)=-2\frac G{N_c}(\bar qq);~~~\pi_a (x)=-2\frac G{N_c}(\bar q
\mathrm{i}\gamma^5\tau_a q).
\label{200}
\end{eqnarray}
Note that the composite bosonic field $\pi_3 (x)$ can be identified
with the physical $\pi_0$ meson, whereas the physical $\pi^\pm
(x)$-meson fields are the following combinations of the composite
fields, 
$\pi^\pm (x)=(\pi_1 (x)\mp i\pi_2 (x))/\sqrt{2}$. 
Obviously, the semi-bosonized Lagrangian $\widetilde L$ is equivalent to the initial Lagrangian (\ref{1}) when using the equations (\ref{200}). Furthermore, it is clear from (\ref{2001}), (\ref{200}) and footnote \ref{f1,1} that the bosonic fields transform under the isospin $U_{I_3}(1)$ and axial isospin $U_{AI_3}(1)$ groups in the following manner:
\begin{eqnarray}
U_{I_3}(1):~&&\sigma\to\sigma;~~\pi_3\to\pi_3;~~\pi_1\to\cos
(\beta)\pi_1+\sin (\beta)\pi_2;~~\pi_2\to\cos (\beta)\pi_2-\sin
(\beta)\pi_1,\nonumber\\
U_{AI_3}(1):~&&\pi_1\to\pi_1;~~\pi_2\to\pi_2;~~\sigma\to\cos
(\omega)\sigma+\sin (\omega)\pi_3;~~\pi_3\to\cos
(\omega)\pi_3-\sin (\omega)\sigma.
\label{201}
\end{eqnarray}
In general the phase structure of a given model is characterized by the behaviour of some quantities, called order parameters (or condensates), vs external conditions (temperature, chemical potentials, etc). In the case of model (1), such order parameters are the ground state expectation values of the composite fields, i.e. the quantities $\vev{\sigma (x)}$ and $\vev{\pi_a (x)}$ $(a=1,2,3)$. It is clear from (\ref{201}) that if $\vev{\sigma(x)}\ne 0$ and/or $\vev{\pi_3(x)}\ne 0$, then the axial isospin $U_{AI_3}(1)$ symmetry of the model is spontaneously broken down, whereas at $\vev{\pi_1(x)}\ne 0$ and/or $\vev{\pi_2(x)}\ne 0$ we have a spontaneous breaking of the isospin $U_{I_3}(1)$ symmetry. Since in the last case the ground state expectation values (condensates) of both the fields $\pi^+(x)$ and $\pi^-(x)$ are not zero, this phase is usually called charged pion condensation (PC) phase. The ground state expectation values $\vev{\sigma(x)}$ and $\vev{\pi_a(x)}$ are the coordinates of the global minimum point of the thermodynamic potential $\Omega (\sigma,\pi_a)$ of the system.

Starting from the linearized semi-bosonized model Lagrangian (\ref{2}), one obtains in the leading order
of the large $N_c$-expansion (i.e. in the one-fermion loop
approximation) the following path integral expression for the
effective action ${\cal S}_{\rm {eff}}(\sigma,\pi_a)$ of the bosonic
$\sigma (x)$ and $\pi_a (x)$ fields:
$$
\exp(\mathrm{i}{\cal S}_{\rm {eff}}(\sigma,\pi_a))=
  N'\int[d\bar q][dq]\exp\Bigl(\mathrm{i}\int\widetilde L\,d^2 x\Bigr),
$$
where
\begin{equation}
{\cal S}_{\rm {eff}}(\sigma,\pi_a)
=-N_c\int d^2x\left [\frac{\sigma^2+\pi^2_a}{4G}
\right ]+\widetilde {\cal S}_{\rm {eff}},
\label{3}
\end{equation}
$N'$ is a normalization constant. The quark contribution to the effective action, i.e. the term $\widetilde {\cal S}_{\rm {eff}}$ in (\ref{3}), is given by:
\begin{equation}
\exp(\mathrm{i}\widetilde {\cal S}_{\rm {eff}})=N'\int [d\bar
q][dq]\exp\Bigl(\mathrm{i}\int\Big [\bar q\mathrm{D}q\Big ]d^4
x\Bigr)=[\Det D]^{N_c}.
 \label{4}
\end{equation}
In (\ref{4}) we have used the notation $\mathrm{D}\equiv D\times
\mathrm{I}_c$, where $\mathrm{I}_c$ is the unit operator in the
$N_c$-dimensional color space and
\begin{equation}
D\equiv\gamma^\rho\mathrm{i}\partial_\rho +\mu\gamma^0+ \nu\tau_3\gamma^0+\nu_{5}\tau_3\gamma^0\gamma^5-\sigma -\mathrm{i}\gamma^5\pi_a\tau_a
\label{5}
\end{equation}
is the Dirac operator, which acts in the flavor, spinor as well as
coordinate spaces only. Using the general formula $\Det D=\exp {\rm
Tr}\ln D$, one obtains for the effective action the following expression
\begin{equation}
{\cal S}_{\rm {eff}}(\sigma,\pi_a)=-N_c\int
d^2x\left[\frac{\sigma^2+\pi^2_a}{4G}\right]-\mathrm{i}N_c{\rm
Tr}_{sfx}\ln D,
\label{6}
\end{equation}
where the Tr-operation stands for the trace in spinor ($s$), flavor
($f$) as well as two-dimensional coordinate ($x$) spaces,
respectively. Using (\ref{6}), we obtain the thermodynamic
potential (TDP) $\Omega (\sigma,\pi_a)$ of the system:
\begin{eqnarray}
\Omega (\sigma,\pi_a)~
&&\equiv -\frac{{\cal S}_{\rm {eff}}(\sigma,\pi_a)}{N_c\int
d^2x}~\bigg |_{~\sigma,\pi_a=\rm {const}}
=\frac{\sigma^2+\pi^2_a}{4G}+\mathrm{i}\frac{{\rm
Tr}_{sfx}\ln D}{\int d^2x}\nonumber\\
&&=\frac{\sigma^2+\pi^2_a}{4G}+\mathrm{i}{\rm
Tr}_{sf}\int\frac{d^2p}{(2\pi)^2}\ln\overline{D}(p),
\label{7}
\end{eqnarray}
where the $\sigma$ and $\pi_a$ fields are now $x$-independent quantities, and
\begin{equation}
\overline{D}(p)=\not\!p +\mu\gamma^0+ \nu\tau_3\gamma^0+ \nu_{5}\tau_3\gamma^0\gamma^5-\sigma
-\mathrm{i}\gamma^5\pi_a\tau_a
\label{50}
\end{equation}
is the momentum space representation of the Dirac operator $D$ (\ref{5}). In what follows we are going to investigate the $\mu,\nu,\nu_{5}$-dependence of the global minimum point of the function $\Omega (\sigma,\pi_a)$ vs $\sigma,\pi_a$. To simplify the task, let us note that due to the $U_{I_3}(1)\times U_{AI_3}(1)$ invariance of the model, the TDP (\ref{7}) depends effectively only on the two combinations $\sigma^2+\pi_3^2$ and $\pi_1^2+\pi_2^2$ of the bosonic fields, as is easily seen from (\ref{201}). In this case, without loss of generality, one can put $\pi_2=\pi_3=0$ in (\ref{7}), and study the TDP (\ref{7}) as a function of only two variables, $M\equiv\sigma$ and $\Delta\equiv\pi_1$. Taking into account this constraint in (\ref{50}) and (\ref{7}) as well as the general relation
\begin{eqnarray}
{\rm Tr}_{sf}\ln\overline{D}(p)=\ln\Det\overline{D}(p)=\sum_i\ln\epsilon_i,
\end{eqnarray}
where the summation over all four eigenvalues $\epsilon_i$ of the 4$\times$4 matrix $\overline{D}(p)$ is implied and
\begin{eqnarray}
\epsilon_{1,2,3,4}=-M\pm\sqrt{(p_0+\mu)^2-p_1^2-\Delta^2+
\nu^2-\nu_{5}^2\pm 2\sqrt{\big [(p_0+\mu)\nu+p_1\nu_{5}\big ]^2-\Delta^2(\nu^2-\nu_{5}^2)}},
\label{8}
\end{eqnarray}
we have from (\ref{7})
\begin{eqnarray}
\Omega (M,\Delta)
=\frac{M^2+\Delta^2}{4G}+\mathrm{i}\int\frac{d^2p}{(2\pi)^2}\ln
P_4(p_0).
\label{9}
\end{eqnarray}
In (\ref{9}) we use the notation
\begin{eqnarray}
P_4(p_0)=\epsilon_1\epsilon_2\epsilon_3\epsilon_4=\eta^4-2a\eta^2-b\eta+c,
\label{91}
\end{eqnarray}
where $\eta=p_0+\mu$ and
\begin{eqnarray}
a&=&M^2+\Delta^2+p_1^2+\nu^2+\nu_{5}^2;~~b=8p_1\nu\nu_{5};\nonumber\\
c&=&a^2-4p_1^2(\nu^2+\nu_5^2)-4M^2\nu^2-4\Delta^2\nu_5^2-4\nu^2\nu_5^2.
\label{10}
\end{eqnarray}
Thus, it follows from Eqs. (\ref{91}) and (\ref{10}) that the TDP (\ref{9}) is invariant with respect to the so-called duality transformation (for an analogous case of duality between chiral and superconducting condensates, see \cite{thies2,ekkz2}),
\begin{eqnarray}
{\cal D}:~~~~M\longleftrightarrow \Delta,~~\nu\longleftrightarrow\nu_5.
 \label{16}
\end{eqnarray}
In powers of $\Delta$ and $M$ the fourth-degree polynomial $P_4(p_0)$ has the following forms
\begin{eqnarray}
P_4(p_0)&\equiv&\Delta^4-2\Delta^2(\eta^2-p_1^2-M^2+\nu_5^2-\nu^2)\nonumber\\
&+&\big [M^2+(p_1-\nu_5)^2-(\eta+\nu)^2\big ]\big [M^2+(p_1+\nu_5)^2-
(\eta-\nu)^2\big ] \label{17}\\
&\equiv& M^4-2M^2(\eta^2-p_1^2-\Delta^2+\nu^2-\nu_5^2)\nonumber\\
&+&\big [\Delta^2+(p_1-\nu)^2-(\eta+\nu_5)^2\big ]\big
[\Delta^2+(p_1+\nu)^2-(\eta-\nu_5)^2\big ].\label{18}
\end{eqnarray}
Note that according to the general theorem of algebra, the polynomial
$P_4(p_0)$ can be presented also in the form
\begin{eqnarray}
P_4(p_0)\equiv (p_0-p_{01})(p_0-p_{02})(p_0-p_{03})(p_0-p_{04}), \label{170}
\end{eqnarray}
where the roots $p_{01}$, $p_{02}$, $p_{03}$ and $p_{04}$ of this polynomial 
are the energies of quasiparticle or quasiantiparticle excitations of the system. In particular, it follows from (\ref{17}) that at $\Delta=0$ the set of roots $p_{0i}$ looks like
\begin{eqnarray}
\Big\{p_{01},p_{02},p_{03},p_{04}\Big\}\Big |_{\Delta=0}=\Big\{-\mu-\nu\pm\sqrt{M^2+(p_1-\nu_5)^2},-\mu+\nu\pm\sqrt{M^2+(p_1+\nu_5)^2}\Big\}, \label{26}
\end{eqnarray}
whereas it is clear from (\ref{18}) that at $M=0$ it has the form
\begin{eqnarray}
\Big\{p_{01},p_{02},p_{03},p_{04}\Big\}\Big |_{M=0}=\Big\{-\mu-\nu_5\pm\sqrt{\Delta^2+(p_1-\nu)^2},-\mu+\nu_5\pm\sqrt{\Delta^2+(p_1+\nu)^2}\Big\}. \label{27}
\end{eqnarray}
Taking into account the relation (\ref{170}) as well as the formula (it is proved, e.g., in Ref. \cite{gkkz})
\begin{eqnarray}
\int_{-\infty}^\infty dp_0\ln\big
(p_0-K)=\mathrm{i}\pi|K|\label{int}
\end{eqnarray}
(being true up to an infinite term independent of the real quantity $K$),
it is possible to integrate in (\ref{9}) over $p_0$. Then, the {\it unrenormalized}
TDP (\ref{9}) can be presented in the following form,
\begin{eqnarray}
\Omega (M,\Delta)&\equiv&\Omega^{un} (M,\Delta)=
\frac{M^2+\Delta^2}{4G}-
\int_{-\infty}^\infty\frac{dp_1}{4\pi}\Big (|p_{01}|+|p_{02}|+|p_{03}|+|p_{04}|\Big ). \label{28}
\end{eqnarray}
It is clear directly from the relations (\ref{91}) and (\ref{10}) that the TDP (\ref{9}) is an even function over each of the variables $M$ and $\Delta$. In addition, it is invariant under each of the transformations
$\mu\to-\mu$,  $\nu\to-\nu$, $\nu_5\to-\nu_5$.
\footnote{Indeed, if we perform simultaneously with $\mu\to-\mu$
the change of variables $p_0\to-p_0$ and $p_1\to-p_1$ in the integral (\ref{9}),
then one can easily see that the expression (\ref{9}) remains
intact. Finally, if only $\nu$ (only $\nu_5$) is replaced by $-\nu$
(by $-\nu_5$), we should transform $p_1\to-p_1$ in the integral (\ref{9}) in order to see that the TDP remains unchanged. } Hence, without loss of generality we can consider in the following only $\mu\ge 0$, $\nu\ge 0$, $\nu_5\ge 0$, $M\ge 0$, and $\Delta\ge 0$ values of these quantities. Finally, note that the integrand in Eq. (\ref{28}) is not an even function vs $p_1$ (see also in Appendix \ref{ApB}). 

\subsection{The nonzero temperature case}
\label{T}

Though, the effect of nonzero temperatures is quite predictable (one can expect that the temperature just restores all the broken symmetries of the model), here we include nonzero temperatures into consideration 
because it is instructive to know how robust the charged PC phase under temperature.

To introduce finite temperature $T$ into consideration, it is very convenient to use the zero temperature expression (\ref{9}) for the TDP,
\begin{eqnarray}
\Omega (M,\Delta)
=\frac{M^2+\Delta^2}{4G}+\mathrm{i}\int\frac{d^2p}{(2\pi)^2}\ln
(p_0-p_{01})(p_0-p_{02})(p_0-p_{03})(p_0-p_{04}),
\label{09}
\end{eqnarray}
where we took into account Eq. (\ref{170}). 
Then, to find the temperature dependent TDP $\Omega_T(M,\Delta)$ one should replace in Eq. (\ref{09}) the integration over $p_0$ in favor of the summation over Matsubara frequencies $\omega_n$ by the rule
\begin{eqnarray}
\int_{-\infty}^{\infty}\frac{dp_0}{2\pi}\big (\cdots\big )\to
iT\sum_{n=-\infty}^{\infty}\big (\cdots\big ),~~~~p_{0}\to
p_{0n}\equiv i\omega_n \equiv i\pi T(2n+1),~~~n=0,\pm 1, \pm 2,...
\label{190}
\end{eqnarray}
In the expression obtained, it is possible to sum over Matsubara frequencies using the general formula (the corresponding technique is presented, e.g., in \cite{jacobs})
\begin{eqnarray}
&&\sum^{\infty}_{n=-\infty}\ln (i\omega_n-a)
=\ln\left [\exp (\beta |a|/2)+\exp (-\beta |a|/2)\right ]
=\frac{\beta
|a|}{2}+\ln\left [1+\exp (-\beta |a|)\right ],\label{C4}
\end{eqnarray}
where $\beta=1/T$. As a result, one can obtain the following expression for the TDP $\Omega_T(M,\Delta)$
\begin{eqnarray}
\Omega_T (M,\Delta)
&=&\Omega (M,\Delta)
-T\sum_{i=1}^{4}\int_{-\infty}^{\infty}\frac{dp_1}{2\pi}\ln\big (1+e^{-\beta|p_{0i}|}\big ),\label{260}
\end{eqnarray}
where $\Omega (M,\Delta)$ is the TDP (\ref{28}) of the system at zero temperature. Since each root $p_{0i}$ in Eq. (\ref{260}) is a dually ${\cal D}$ invariant quantity (see in Appendix \ref{ApB}), it is clear that the temperature dependent TDP (\ref{260}) is also symmetric with respect to the duality transformation ${\cal D}$ (\ref{16}).

\section{Calculation of the TDP at $T=0$}
\subsection{Thermodynamic potential in vacuum: the case of $\mu=0, \nu=0,\nu_5=0$}

First of all, let us obtain a finite, i.e. renormalized, expression
for the TDP (\ref{28}) at $\mu=0$, $\nu=0$ and $\nu_5=0$, i.e. in
vacuum, and at zero temperature. Since in this case a thermodynamic potential is usually called effective potential, we use for it the notation $V^{un} (M,\Delta)$. As a consequence of (\ref{9})-(\ref{10}) and using (\ref{int}),
it is clear that at $\mu=\nu=\nu_5=0$ the effective potential $V^{un} (M,\Delta)$ looks like
\begin{eqnarray}
V^{un} (M,\Delta)&=&
\frac{M^2+\Delta^2}{4G}
+2i\int\frac{d^2p}{(2\pi)^2}\ln\Big
[p_0^2-p_1^2-M^2-\Delta^2\Big ]\nonumber\\
&=&\frac{M^2+\Delta^2}{4G}-\int_{-\infty}^\infty\frac{dp_1}{\pi}\sqrt{p_1^2+M^2+\Delta^2}. \label{25}
\end{eqnarray}
It is evident that the effective potential (\ref{25}) is an ultraviolet (UV) divergent
quantity. So, we need to renormalize it. This procedure consists of
two steps: (i) First of all we need to regularize the divergent integral in (\ref{25}), i.e. we suppose there that $|p_1|<\Lambda$. (ii)  Second, we must suppose also that the bare coupling constant $G$ depends on the cutoff parameter
$\Lambda$ in such a way that in the limit $\Lambda\to\infty$ one obtains a finite expression for the effective potential.

Following the step (i) of this procedure, we have
\begin{eqnarray}
V^{reg} (M,\Delta)&=&\frac{M^2+\Delta^2}{4G}-\frac 2\pi\int_{0}^\Lambda dp_1\sqrt{p_1^2+M^2+\Delta^2}\nonumber\\
=\frac{M^2+\Delta^2}{4G}&-&\frac 1\pi\left\{\Lambda\sqrt{\Lambda^2+M^2+\Delta^2}+(M^2+\Delta^2)\ln\frac{\Lambda+\sqrt{\Lambda^2+M^2+\Delta^2}}{\sqrt{M^2+\Delta^2}}\right\}. \label{29}
\end{eqnarray}
Further, according to the step (ii) we suppose that in (\ref{29}) the bare coupling $G\equiv G(\Lambda)$ has the following $\Lambda$ dependence:
\begin{eqnarray}
\frac 1{4G(\Lambda)}=\frac 1\pi\ln\frac{2\Lambda}{m}, \label{30}
\end{eqnarray}
where $m$ is a new mass scale of the model, and $m$ is a free model parameter. It appears instead of the dimensionless bare 
coupling constant $G$ (dimensional transmutation) and, evidently, does not depend on a normalization point, i.e. it is a renormalization
invariant quantity. Substituting (\ref{30}) into (\ref{29}) and
ignoring there an unessential term $(-\Lambda^2/\pi)$, we obtain
in the limit $\Lambda\to\infty$ the finite and renormalization invariant expression for the effective potential,
\begin{eqnarray}
V_0 (M,\Delta)&=&\frac{M^2+\Delta^2}{2\pi}\left [\ln\left (\frac{M^2+\Delta^2}{m^2}\right )-1\right ]. \label{31}
\end{eqnarray}

\subsection{Calculation of the TDP (\ref{28}) in the general case: $\mu>0$, $\nu>0$, $\nu_5>0$}\label{IIIB}

In Appendix \ref{ApB} the properties of the quasiparticle energies
$p_{0i}$, where $i=1,...,4$, are investigated. In particular, it is
clear from the asymptotic expansion (\ref{B9}) that the integral
over $p_1$ in (\ref{28}) is ultraviolet divergent. It is possible to transform the expression (\ref{28}) in the following way,
\begin{eqnarray}
\Omega^{un} (M,\Delta)&=&
\frac{M^2+\Delta^2}{4G}-
\int_{-\infty}^\infty\frac{dp_1}{4\pi}\Big (|p_{01}|+|p_{02}|+|p_{03}|+|p_{04}|\Big )\Big |_{\mu=\nu=\nu_5=0}\nonumber\\
&-&\int_{-\infty}^\infty\frac{dp_1}{4\pi}\Big [\sum_{i=1}^4|p_{0i}|-\Big (\sum_{i=1}^4|p_{0i}|\Big )\Big |_{\mu=\nu=\nu_5=0}\Big ]. \label{32}
\end{eqnarray}
Since the asymptotic expansion (\ref{B9}) does not depend on chemical potentials $\mu$, $\nu$ and $\nu_5$, it is
evident that the last integral in (\ref{32}) is a convergent and all UV divergences of the TDP are located in the first integral of
(\ref{32}). Moreover, it is clear due to the relation (\ref{B10}) that
the first two terms in the right hand side of Eq. (\ref{32}) are just
the unrenormalized effective potential in vacuum (\ref{25}). So to
obtain a finite expression for the TDP (\ref{32}), it is enough to proceed as in the
previous subsection, where just these two terms, i.e. the vacuum
effective potential, were renormalized. As a result, we have
\begin{eqnarray}
\Omega^{ren}
(M,\Delta)&=&V_0(M,\Delta)-\int^\infty_{-\infty}\frac{dp_1}{4\pi}\Big\{|p_{01}|+|p_{02}|+|p_{03}|+|p_{04}|-4\sqrt{p_1^2+M^2+\Delta^2}\Big\}, \label{35}
\end{eqnarray}
where $V_0(M,\Delta)$ is the renormalized TDP (effective potential) (\ref{31}) of the model at $\mu=\nu=\mu_5=0$. Moreover, we have used in (\ref{35}) the relation (\ref{B10}) for the sum of quasiparticle energies in vacuum. Note also that (as it follows from the considerations of Appendix A) the quasiparticle energies $p_{0i}$, where $i=1,...,4$, are invariant 
with respect to the duality transformation (\ref{16}). So the renormalized TDP (\ref{35}) is also symmetric under the duality transformation $\cal D$. 

In Ref. \cite{ekk} the erroneous assumption was used that the quantity $\sum_{i=1}^4|p_{0i}|$ is an even function vs $p_1$. 
This further led to an incorrect expression for the TDP (\ref{35}), in which the integration over $p_1\in(0,+\infty)$ was used, 
and to erroneous phase diagrams. Here we correct this mistake. Therefore, in Eq. (\ref{35}), the integration over $p_1\in(-\infty,+\infty)$ is carried out. However, the main qualitative conclusions about the properties of the chiral isospin chemical potential $\nu_5$ and the duality of the model under consideration do not change.

Let us denote by $(M_0,\Delta_0)$ the global minimum point (GMP) of the TDP (\ref{35}). Then, investigating the behavior of this 
point vs $\mu$, $\nu$ and $\nu_5$ it is possible to construct the $(\mu,\nu,\nu_5)$-phase portrait (diagram) of the model. 
A numerical algorithm for finding the quasi(anti)particle energies  $p_{01}$, $p_{02}$, $p_{03}$, and $p_{04}$ is elaborated in 
Appendix \ref{ApB}. Based on this algorithm, it is possible to study the TDP of the model in the region of sufficiently small 
values of chemical potentials, e.g., at $\mu,\nu,\nu_{5}\le 2m$, where we did not find any GMP of the form 
$(M_0\ne 0,\Delta_0\ne 0)$ (this conclusion probably will not change at larger values of chemical potentials). Strictly speaking, 
this property of the TDP is typical for the NJL models only in the chiral limit. At the physical point, i.e. when quarks have a 
nonzero bare mass, in the charged PC phase both $\Delta_0$ and $M_0$ condensates are nonzero (see, e.g., in Refs \cite{he,kkz2,adhikari}).
But these condensates are of the same order of magnitude only in the transitory 
region (in the region between two phases, CSB and charged PC), whereas deeper in the phases 
one of the condensates dominates. For example, in the charged PC phase, first (at $\nu$ slightly larger than half of pion mass) 
there is a non-zero chiral condensate, or dynamical quark mass, $M_0\gg\Delta_0$ 
but it decreases very fast and at larger $\nu$ chiral condensate $M_0$ is much smaller than the charged pion one, $\Delta_0$. 
Let us also recall that chiral limit overall is a good approximation as it has been shown in \cite{kkz2}.
Hence, in order to establish the phase portrait of the massless model (1), e.g, in the region $\mu,\nu,\nu_{5}\le 2m$, it is enough to study the projections 
$F_1(M)\equiv\Omega^{ren} (M,\Delta=0)$ and $F_2(\Delta)\equiv\Omega^{ren}(M=0,\Delta)$ of 
the TDP (\ref{35}) to the $M$ and $\Delta$ axes, correspondingly.  Taking into account the relations (\ref{26}) 
for the quasiparticle energies $p_{0i}$  at $\Delta=0$, 
it is possible  to obtain the following expression for the projection $F_1(M)$,
\begin{eqnarray}
F_1(M)&=&\frac{M^2}{2\pi}\ln\left
(\frac{M^2}{m^2}\right )-\frac{M^2}{2\pi}-\frac{\nu_5^2}{\pi}\nonumber\\
&-&\frac{\theta (|\mu-\nu|-M)}{2\pi}\left
(|\mu-\nu|\sqrt{(\mu-\nu)^2-M^2}
-M^2\ln\frac{|\mu-\nu|+\sqrt{|\mu-\nu|^2-M^2}}{M}\right )\nonumber\\
&-&\frac{\theta (\mu+\nu-M)}{2\pi}\left ((\mu+\nu)\sqrt{(\mu+\nu)^2-M^2}
-M^2\ln\frac{\mu+\nu+\sqrt{(\mu+\nu)^2-M^2}}{M}\right ).\label{33}
\end{eqnarray}
Then, the projection $F_2(\Delta)$ of the TDP on the $\Delta$ axis can be obtained directly from Eq. (\ref{27}) or, alternatively,
using the dual symmetry ${\cal D}$ of the TDP,
\begin{eqnarray}
F_2(\Delta)=F_1(\Delta)\Bigg |_{\nu\longleftrightarrow\nu_5}. \label{34}
\end{eqnarray}
(Details of the derivation of these expressions are given in Appendix
\ref{ApD}.) After simple transformations, one can see that $F_1(M)$ and $F_2(\Delta)$ coincide at $\nu_5=0$ with 
corresponding TDPs (12) and (13) of the paper \cite{ek2}.
Moreover, it is obvious that the GMP of the TDP (\ref{35}) is defined by a comparison between the least values of the functions $F_1(M)$ and $F_2(\Delta)$.

\subsection{Quark number density}

As it is clear from the above consideration, there are three phases in
the model (1).  The first one is the symmetric phase, which
corresponds to the global minimum point $(M_0,\Delta_0)$ of the TDP
(\ref{35}) of the form $(M_0=0,\Delta_0=0)$. In the CSB phase the TDP
reaches the least value at the point $(M_0\ne 0,\Delta_0=0)$. Finally,
in the charged PC phase the GMP lies at the point $(M_0=0,\Delta_0\ne 0)$. (Notice, that in the most general case the coordinates (condensates) $M_0$ and $\Delta_0$ of the global minimum point depend on chemical potentials.)

In the present subsection we would like to obtain the expression for the quark number (or particle) density $n_q$ in the ground state of each phase. Recall that in the most general case this quantity is defined by the relation \footnote{The density of baryons $n_B$ and the quark number density $n_q$ are connected by the relation $n_q=3n_B$.}
\begin{eqnarray}
n_q=-\frac{\partial\Omega^{ren}(M_0,\Delta_0)}{\partial\mu}. \label{37}
\end{eqnarray}
Hence, in the chiral symmetry breaking phase we have
\begin{eqnarray}
n_q\bigg |_{CSB}&=&-\frac{\partial\Omega^{ren}(M_0\ne 0,\Delta_0=0)}{\partial\mu}=-\frac{\partial F_1(M_0)}{\partial\mu}=\frac{\theta\left (\mu+\nu-M_0\right )}{\pi}\sqrt{(\mu+\nu)^2-M_0^2}\nonumber\\
&+&\frac{{\rm sign}(\mu-\nu)\theta\left (|\mu-\nu|-M_0\right )}{\pi}\sqrt{(\mu-\nu)^2-M_0^2}, \label{38}
\end{eqnarray}
where ${\rm sign(x)}$ denotes the sign function and the quantity $F_1(M)$ is given in (\ref{33}).
The quark number density in the charged pion condensation phase can be
easily obtained from (\ref{38}) by the simple replacement,
\begin{eqnarray}
n_q\bigg |_{PC}=-\frac{\partial\Omega^{ren}(M_0=0,\Delta_0\ne 0)}{\partial\mu}=-\frac{\partial F_2(\Delta_0)}{\partial\mu}=\left\{n_q\big |_{CSB}\right\}\bigg |_{M_0\to\Delta_0;~\nu\longleftrightarrow\nu_5}, \label{39}\end{eqnarray}
which is due to the relation (\ref{34}). Supposing in (\ref{38}) that $M_0=0$, one can find the following expression for the 
particle density in the symmetric phase (of course, we take into account the constraints $\mu\ge 0$, $\nu\ge 0$ and $\nu_5\ge 0$)
\begin{eqnarray}
n_q\bigg |_{SYM}&=&\frac{\mu+\nu}{\pi}
+\frac{{\rm sign}(\mu-\nu)}{\pi}|\mu-\nu|=\frac{2\mu}{\pi}. \label{40}
\end{eqnarray}

\section{Phase structure at zero temperature}
\subsection{The role of the duality symmmetry ${\cal D}$ (\ref{16}) of the TDP}

Suppose now that at some fixed particular values of chemical potentials $\mu$, $\nu=A$ and $\nu_5=B$ the global minimum of the TDP (\ref{35}) lies at the
point, e.g., $(M=M_0\ne 0,\Delta=0)$. It means that for such fixed values
of the chemical potentials the CSB phase is realized in the model. Then it follows from the duality invariance of the
TDP (\ref{9}) (or (\ref{35})) with respect to the transformation ${\cal D}$
(\ref{16}) that the permutation of the chemical potential values
(i.e. $\nu=B$ and $\nu_5=A$ and intact value of $\mu$) moves
the global minimum of the TDP $\Omega^{ren}(M,\Delta)$ to the point
$(M=0,\Delta=M_0)$, which corresponds to the charged PC phase
(and vice versa). This is the so-called duality correspondence
between CSB and charged PC phases in the framework of the model under
consideration. \footnote{It is worth to note that in some (1+1)- and (2+1)-dimensional models there is a duality between CSB and superconductivity \cite{thies2,ekkz2}.}

Hence, the knowledge of a phase of the model (1) at
some fixed values of external free model parameters
$\mu,\nu,\nu_5$ is sufficient  to understand what phase (we call it a dually conjugated phase) is realized at rearranged values of external parameters,
$\nu\leftrightarrow\nu_5$, at fixed $\mu$. Moreover, different physical parameters such as condensates, densities, etc, which characterize both the
initial phase and the dually conjugated phase, are connected by the duality transformation ${\cal D}$. For example, the chiral condensate of the initial CSB phase at some fixed $\mu,\nu,\nu_5$ is equal to the charged-pion condensate of the dually conjugated charged PC phase, in which one should perform the replacement  $\nu\leftrightarrow\nu_5$. Knowing the particle density $n_q$ of the initial CSB phase (see in Eq. (\ref{38})) as a function of external chemical potentials $\mu$, $\nu$ and $\nu_{5}$, one can find the particle density (\ref{39}) in the dually conjugated charged PC phase by interchanging $\nu$ and $\nu_{5}$ in the initial expression $n_q$ for the particle density in the CSB phase (see also in Sec. III C), etc.

The duality transformation ${\cal D}$ of the  TDP can also be
applied to an arbitrary phase portrait of the model (see below). In
particular, it is clear that if we have a most general phase
portrait, i.e. the correspondence between any point
$(\nu,\nu_5,\mu)$ of the three-dimensional space of external
parameters and possible model phases (CSB, charged PC and symmetric phase),
then under the duality transformation ($\nu\leftrightarrow\nu_5$, CSB$\leftrightarrow$charged PC and simmetrical phase does not change) this phase portrait is mapped to itself, i.e. the most general $(\nu,\nu_5,\mu)$-phase portrait
is self-dual. The self-duality of the $(\nu,\nu_5,\mu)$-phase portrait
means that the regions of the CSB and charged PC phases in the
three-dimensional $(\nu,\nu_5,\mu)$ space are arranged mirror
symmetrically with respect to the plane $\nu=\nu_5$ of this
space. In the next subsection \ref{VB}, we will present a few sections of this three-dimensional $(\nu,\nu_5,\mu)$-phase portrait of the model by the planes of the form $\mu=const$, $\nu=const$ and $\nu_5=const$, respectively.
\begin{figure}
\includegraphics[width=0.323\textwidth]{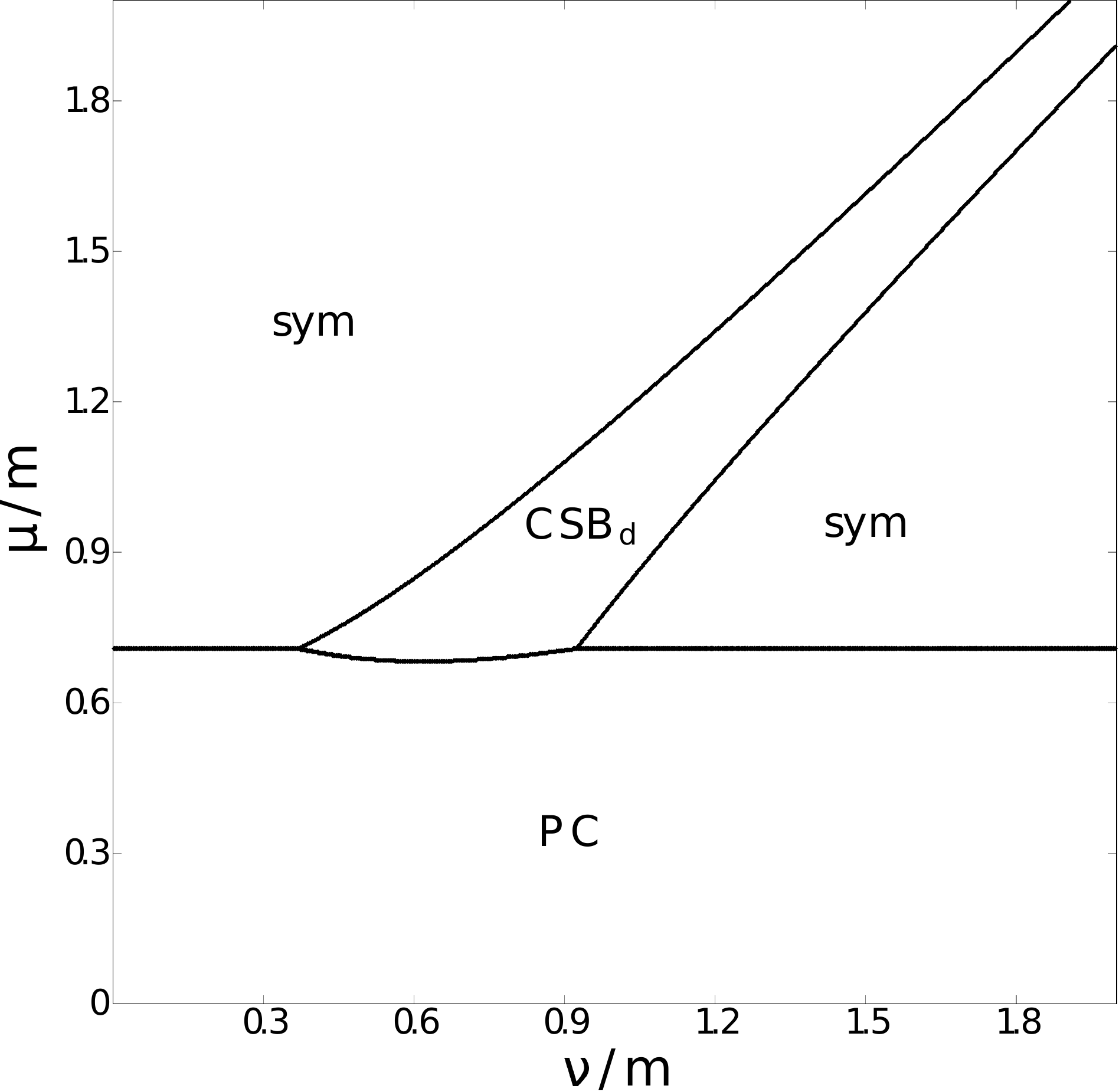}
\includegraphics[width=0.323\textwidth]{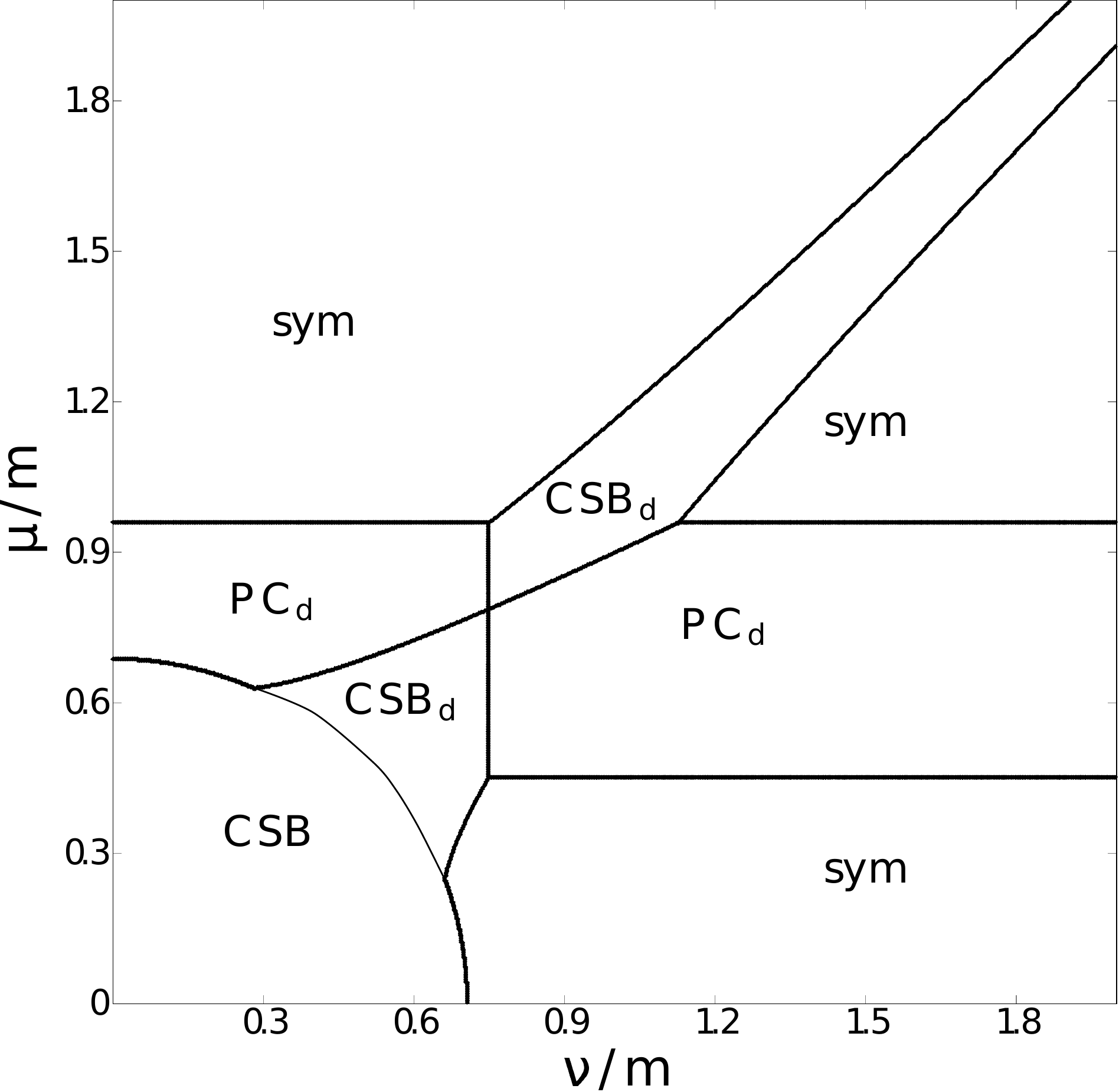}
\includegraphics[width=0.323\textwidth]{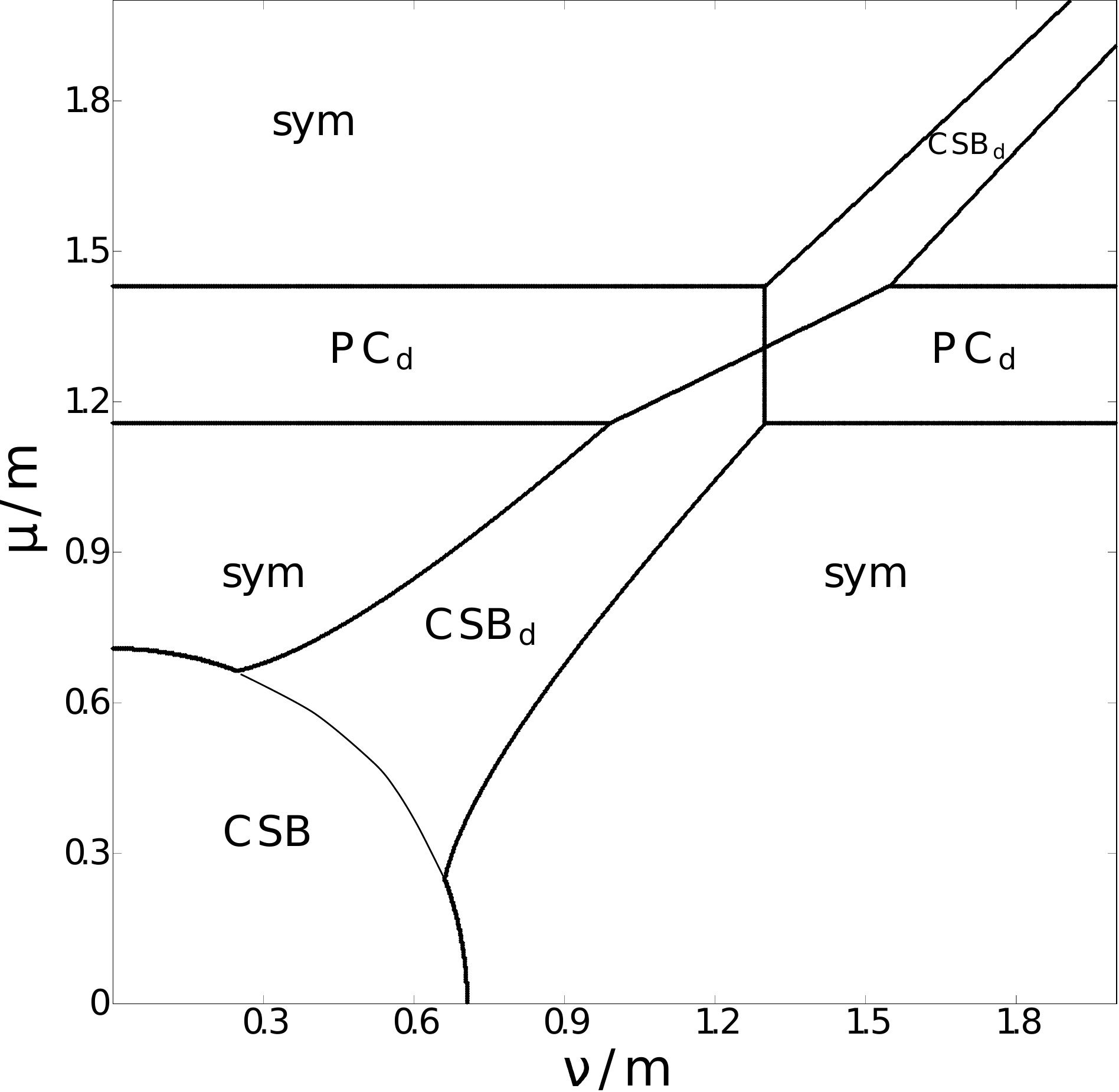}
\caption{The $(\nu,\mu)$-phase portraits of the model for different
values of the chiral chemical potential $\nu_5$: the case
$\nu_5=0$ (left panel), the case  $\nu_5=0.75m$ (middle panel) and the case $\nu_5=1.3m$ (right panel). The notations PC and PC$_d$ mean
the charged pion condensation phase with zero and nonzero baryon
density, respectively. Analogously, the notations CSB and CSB$_d$ mean the chiral symmetry breaking phase with zero and nonzero baryon density, respectively,
and SYM denotes the symmetric phase. The parameter $m$ is introduced in (\ref{30}). 
All the lines are solid and denote phase transitions of the first order.}
\end{figure}

\subsection{Promotion of dense charged PC phase by $\nu_5\ne 0$}
\label{VB}

First of all, we will study the phase structure of the model (1) at different fixed values of the chiral isospin chemical potential $\nu_5$. To this end, we determine numerically the global minimum points of the TDPs $F_1(M)$ (\ref{33})
and $F_2(\Delta)$ (\ref{34}) and then compare the minimum values of
these functions vs external parameters $\mu,\nu,\nu_5$. Moreover,
using the expressions (\ref{38}) and (\ref{39}), it is possible to find the
quark number density $n_q$ or baryon density $n_B$ (note that
$n_q=3n_B$) inside each phase. As a result, in Fig. 1 we have
drawn several $(\nu,\mu)$-phase portraits, corresponding to 
$\nu_5=0$, $\nu_5=0.75m$ and $\nu_5=1.3m$, respectively. Recall that $m$ is a free renormalization invariant mass scale
parameter, which appeares in the vacuum case of the model after renormalization (see Eqs. (\ref{30}) and (\ref{31})).

The phase portrait of the model in Fig. 1 with $\nu_5=0$ was obtained earlier (see e.g. papers \cite{ekkz,ek2}). It is clear from it that at
$\nu_5=0$ the charged PC phase with {\it nonzero baryon density} $n_B$
is not realized in the model under consideration. Only the charged PC phase with {\it zero baryon density} can be observed at rather 
small values of $\mu$. (Physically, it means that at $\nu_5=0$ the model predicts the charged PC phenomenon in the medium 
with $n_B=0$ only. For example, it might consist of charged pions, etc. But in quark matter with nonzero baryon density the 
charged PC is forbidden.) Instead, at large values of $\mu$ there exist two phases, the chiral symmetry breaking and the symmetrical 
one, both with nonzero baryon density, i.e. the model predicts only the CSB or symmetrical phases of dense quark matter at $\nu_5=0$.
However, as we can see from other phase diagrams of Fig. 1, at rather high values of $\nu_5$, e.g., at $\nu_5=0.75m$ or $\nu_5=1.3m$,
there might appear on the phase portrait a charged PC phase with {\it nonzero baryon density} (in Fig. 1 and other figures it is 
denoted as PC$_d$). Hence, in chirally asymmetric, i.e. at $\nu_5>0$, and dense quark matter the charged PC phenomenon is allowed to exist in the framework of the toy model (1). Thus, we see that $\nu_5\ne 0$ is a factor which promotes the charged PC phenomenon in dense quark matter. 

Now, suppose that we want to obtain a $(\nu_5,\mu)$-phase portrait of
the model at some fixed value $\nu=const$. In this case there is no
need to perform the direct numerical investigations of the TDP
(\ref{35}). In contrast (due to the dual invariance (\ref{16}) of the
model TDP), one can simply make the dual transformation of the
$(\nu,\mu)$-phase diagram at the corresponding fixed value
$\nu_5=const$. For example, to find the $(\nu_5,\mu)$-phase diagram at
$\nu=0$ we should start from the $(\nu,\mu)$-diagram at fixed
$\nu_5=0$ of Fig. 1 and make the simplest replacement in the
notations of this figure: $\nu\to\nu_5$, PC$\leftrightarrow$CSB,
PCd$\leftrightarrow$CSBd and the notation ``sym'' does not change. As a result of this mapping, we obtain the phase 
diagram of Fig. 2 (left panel) with PC$_d$ phase. From this phase diagram, we can draw another interesting conclusion that 
the PC$_d$ phase can occur in a dense, {\it isotopically symmetric} baryonic medium, $\nu=0$, but at the same time $\nu_5$ should be non-zero.

In a similar way, to obtain the $(\nu_5,\mu)$-phase diagram, e.g., at $\nu=1.3m$, it is sufficient to apply the duality transformation to 
the $(\nu,\mu)$-phase portrait of the model at $\nu_5=1.3m$ (see the right panel of Fig. 1). The resulting mapping is the right panel of Fig. 2, etc. It thus supports the above conclusion: the charged PC phenomenon can be realized in chirally asymmetric quark matter with nonzero baryon density.
\begin{figure}
\includegraphics[width=0.45\textwidth]{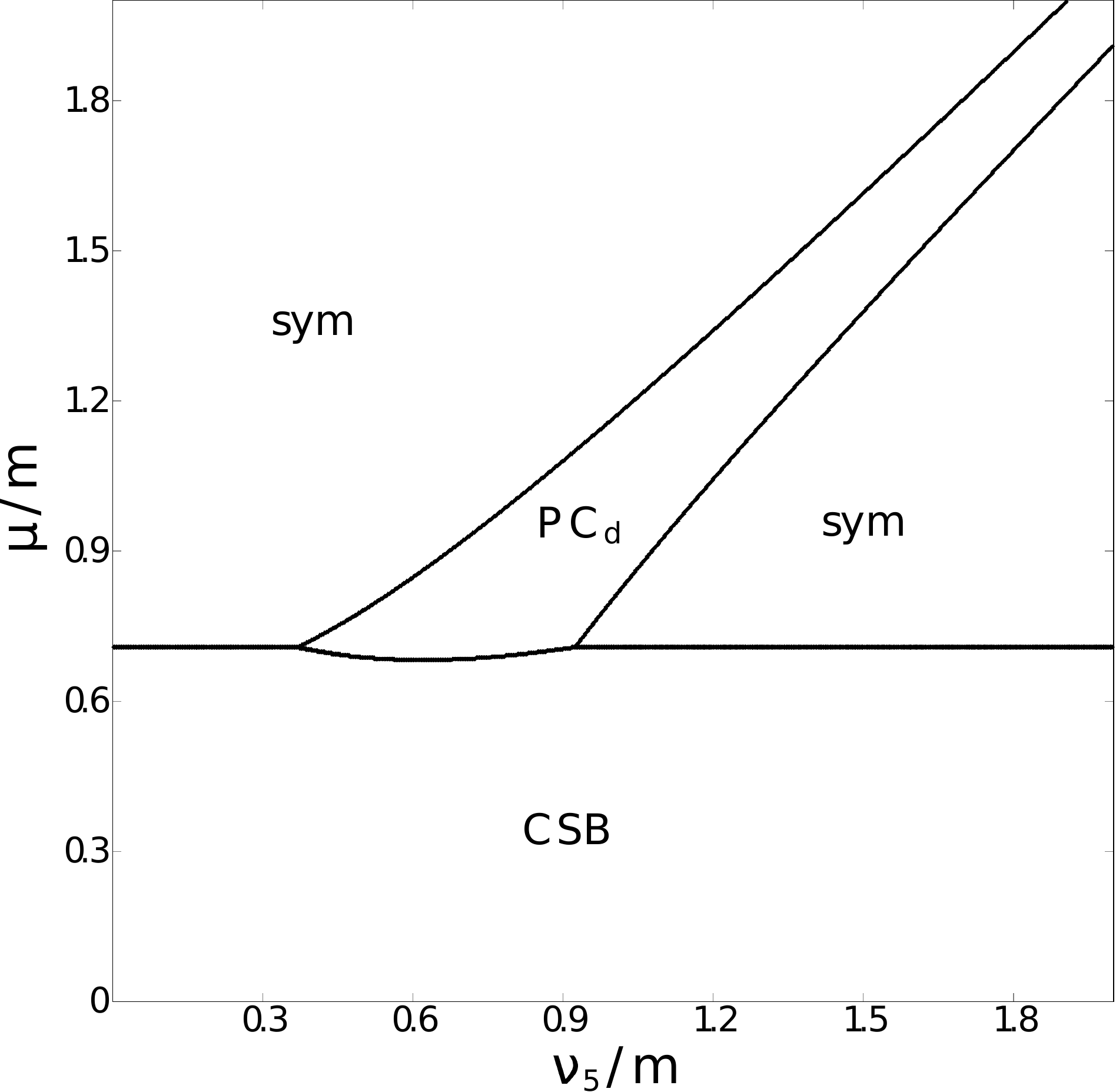}
\hfill
\includegraphics[width=0.45\textwidth]{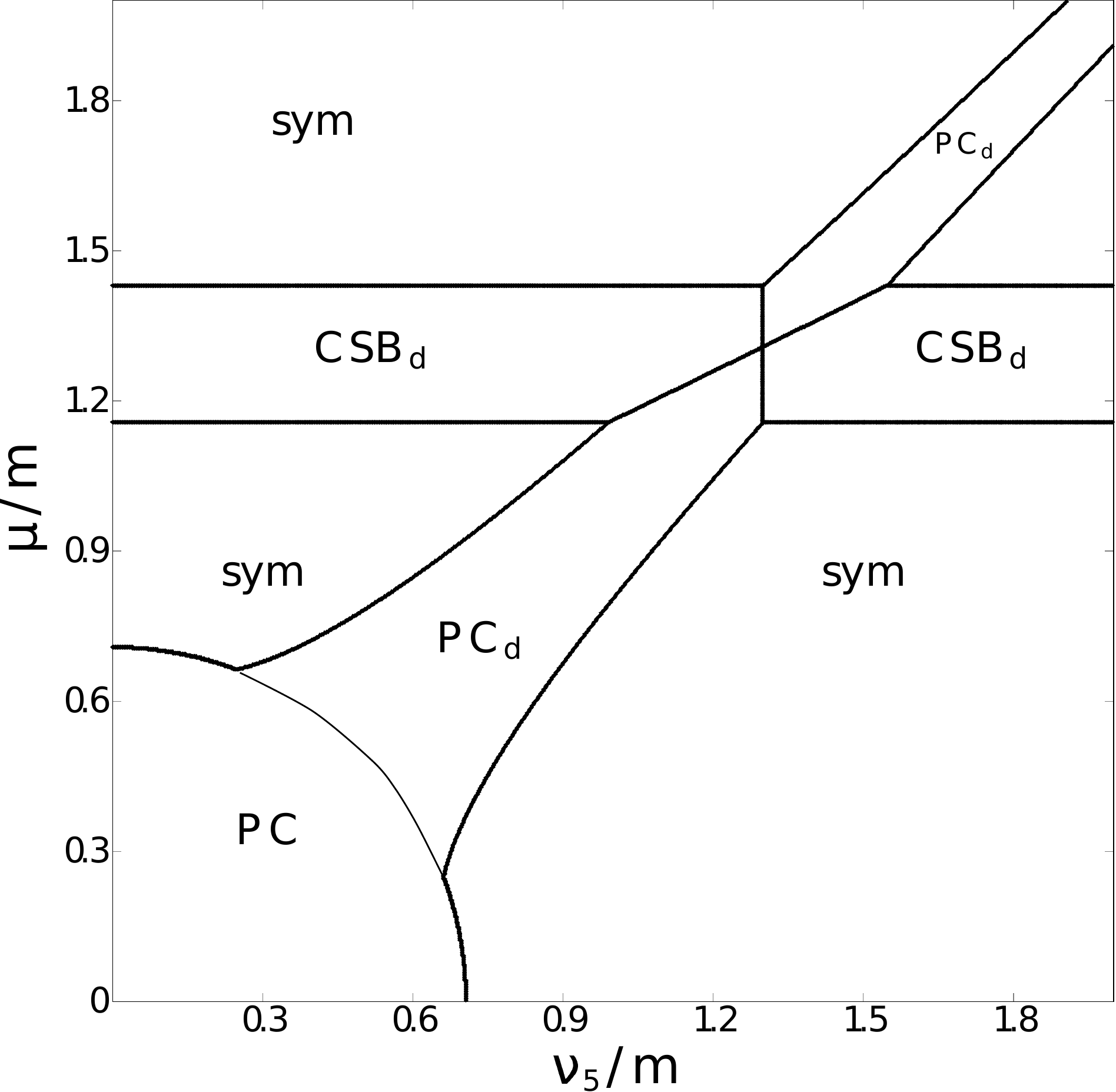}
\caption{The $(\nu_5,\mu)$-phase portrait of the model for different
values of the isospin chemical potential $\nu$: the case
$\nu=0$ (left panel) and case  $\nu=1.3m$ (right panel). All the lines are the first order phase transition curves. The notations are the same as in Fig. 1.}
\end{figure}

Finally, let us consider the $(\nu,\nu_5)$-phase diagrams of the model
at different fixed values of $\mu$. It is clear from the previous
discussions that each of these diagrams is a self-dual one, i.e. the CSB and charged PC phases are arranged symmetrically with respect to the line $\nu=\nu_5$ of the $(\nu,\nu_5)$-plane. This fact is confirmed by $(\nu,\nu_5)$-phase portraits of Fig. 3, obtained by direct numerical analysis of the TDPs $F_1(M)$ (\ref{33}) and $F_2(\Delta)$ (\ref{34}). Moreover, the phase diagrams of Fig. 3 support once again 
the main conclusion of our paper: the charged PC phase with nonzero baryon density, i.e. the phase denoted in Figs 1--3 as PC$_d$,
might be realized in the framework of the model (1) only at $\nu_5\ne 0$.

We have already noted that the phase structure of this NJL$_2$ model has been previously studied in Ref. \cite{ekk}. And 
there the conclusion was drawn that the chiral isospin chemical potential $\mu_{I5}\equiv 2\nu_{5}$ generates the charged 
PC phase in dense quark matter, but only for sufficiently small values of the isospin chemical potential 
$\mu_I\equiv 2\nu$. However, this result was significantly different from the predictions of the NJL$_4$ model, which 
were obtained later in Refs. \cite{kkz,kkz2}, where it was stated that charged PC phase can exist in chirally asymmetric 
medium only at sufficiently high values of the isospin chemical potential $\mu_I$. Based on this 
latter result, it is possible to 
predict that the charged PC phase can be realized in the cores of such astrophysical objects as neutron stars, in which 
both isotopic $\mu_I$ and chiral isotopic $\mu_{I5}$ chemical potential can reach large values. (The possibility of the 
appearance of regions with $\mu_{I5}\ne 0$ inside neutron stars was discussed in Appendix A to the paper \cite{kkz2}). 
In this case, the charged PC phenomenon can have a significant influence on the physical processes in neutron stars and 
lead to observable effects. However, as noted in the Introduction, the results of Ref. \cite{ekk} are based on an 
erroneous expression for the TDP. The correct phase diagrams of the NJL$_2$ model (1) are presented just in the present 
work, from which it can be seen (see Figs. 1 and 3) that the charged PC$_d$ phase occupies an infinitely long strip along 
the $\nu$ axis (for some fixed values of $\nu_5$), which is largely consistent with the predictions of the NJL$_4$ model 
(see in Refs. \cite{kkz,kkz2}). 
Thus, we see that two generally different effective models, NJL$_2$ and NJL$_4$, predict qualitatively identical properties of quark matter with chiral isotopic asymmetry, which, in our opinion, enhances the realism of their predictions.

\begin{figure}
\includegraphics[width=0.45\textwidth]{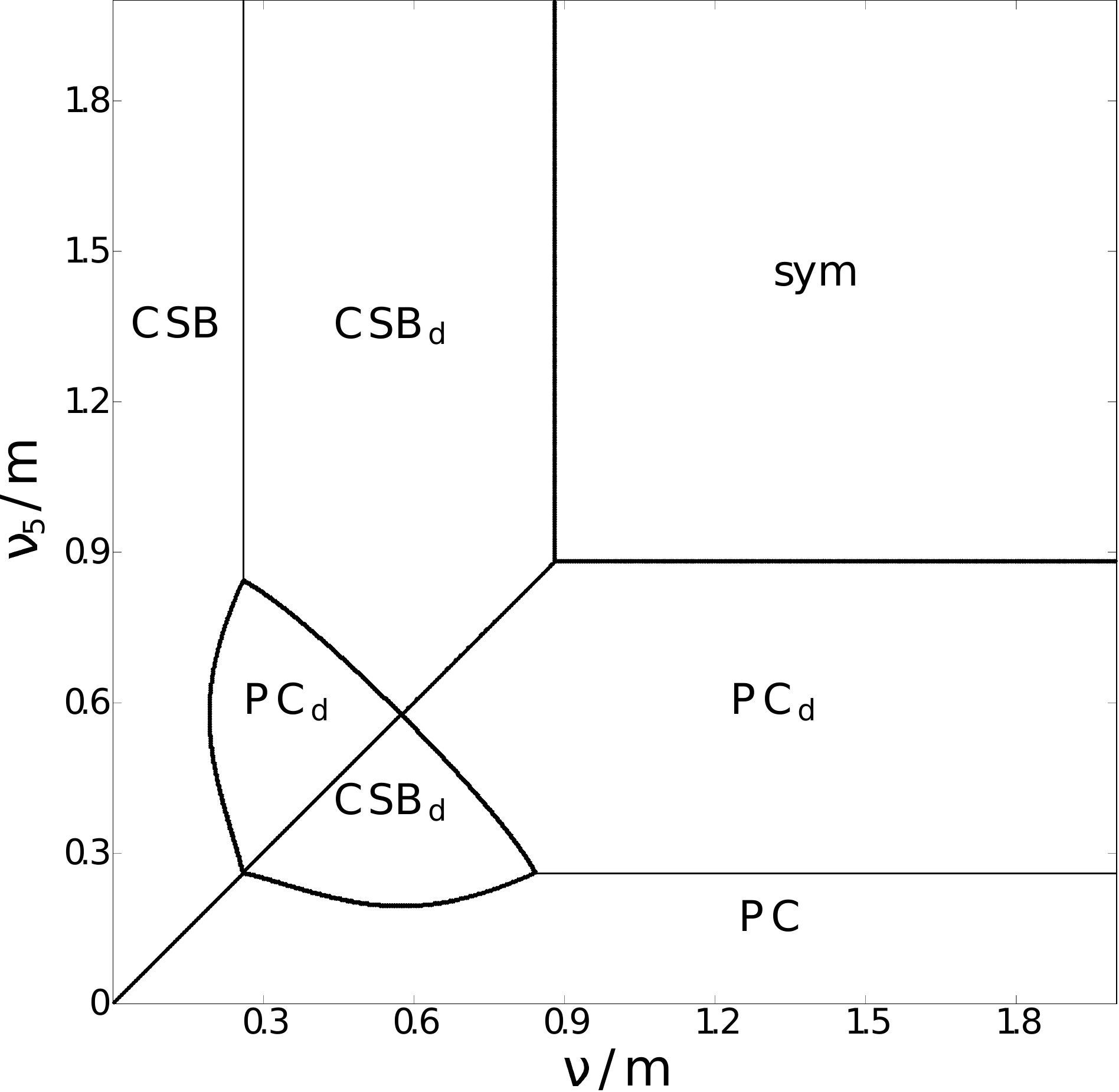}
\hfill
\includegraphics[width=0.45\textwidth]{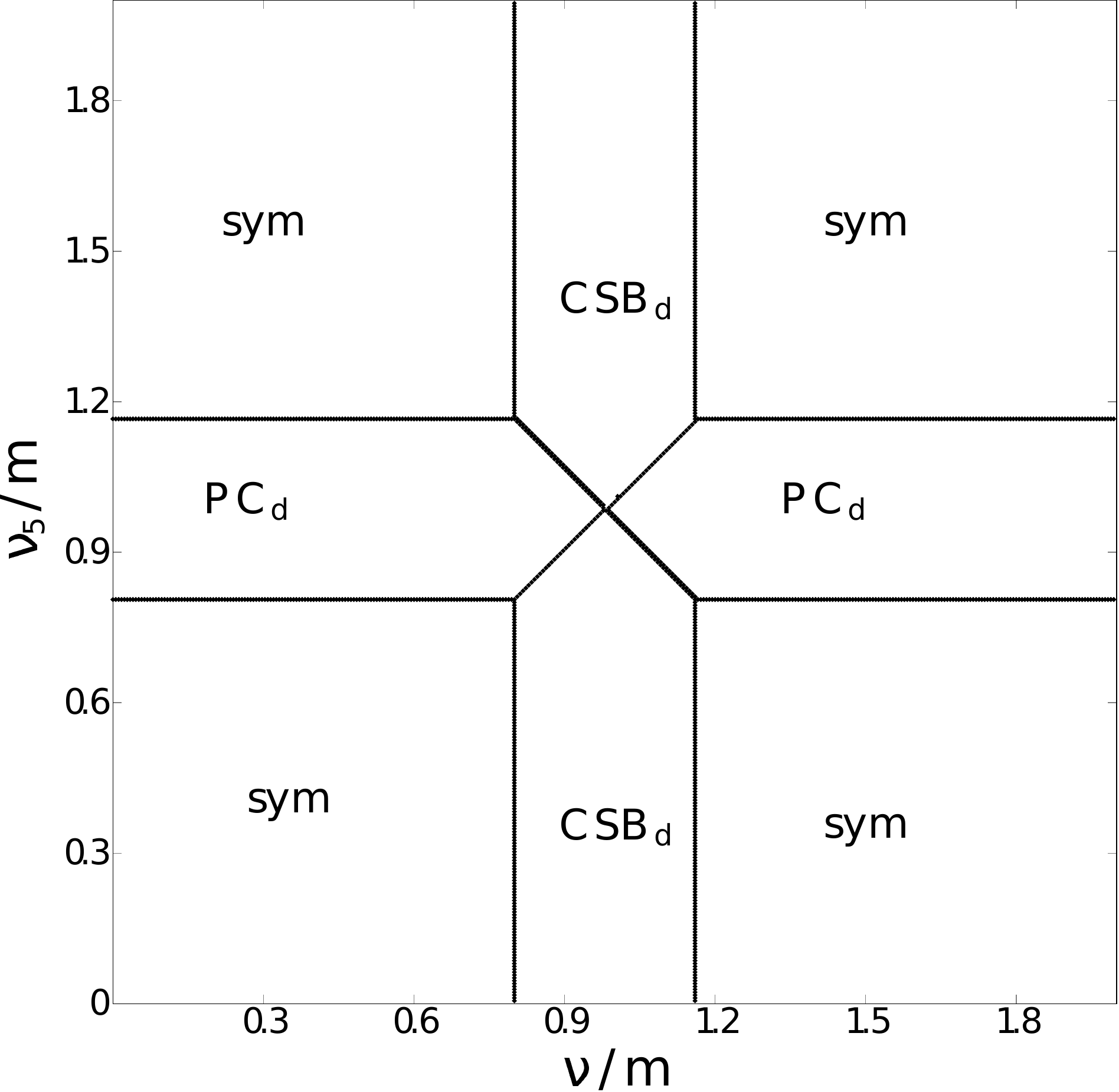}
\caption{The $(\nu,\nu_5)$-phase portraits of the model for different
values of the  quark number chemical potential $\mu$: the case
$\mu=0.65m$ (left panel) and the case  $\mu=m$ (right panel). All the lines are the first order phase transition curves. The notations are the same as in Fig. 1.}
\end{figure}

\section{Phase structure at $T\ne 0$}

First of all note that the temperature dependent TDP $\Omega_T (M,\Delta)$ in Eq. (\ref{260}) is an UV-divergent quantity. While the second term on 
the right side of this relation is a convergent improper integral that does not contain UV divergences, the first term 
$\Omega (M,\Delta)$ is an UV-divergent TDP of the model at $T=0$. So both the TDP $\Omega (M,\Delta)$  at $T=0$ and the 
TDP $\Omega_T (M,\Delta)$ at nonzero temperature can be renormalized using the technique presented in the section \ref{IIIB}. As a 
result, one can obtain the following expression for the renormalized TDP of the model at $T\ne 0$
\begin{eqnarray}
\Omega_T^{ren} (M,\Delta)
&=&\Omega^{ren} (M,\Delta)
-T\sum_{i=1}^{4}\int_{-\infty}^{\infty}\frac{dp_1}{2\pi}\ln\big (1+e^{-\beta|p_{0i}|}\big ),\label{026}
\end{eqnarray}
where $\Omega^{ren} (M,\Delta)$ is presented in Eq. (\ref{35}). A numerical algorithm for finding the quasi(anti)particle 
energies  $p_{01}$, $p_{02}$, $p_{03}$, and $p_{04}$ is elaborated in Appendix \ref{ApB}. Based on this, it can be shown numerically that the
GMP of the TDP (\ref{026}) can never be of the form $(M_0\ne 0,\Delta_0\ne 0)$. Hence, in order to establish the phase portrait of the model at $T\ne 0$, it is enough to study the projections $F_1^T(M)\equiv\Omega_T^{ren} (M,\Delta=0)$ and $F_2^T(\Delta)\equiv\Omega_T^{ren}(M=0,\Delta)$ of the TDP (\ref{026}) to the $M$ and $\Delta$ axes, correspondingly. Taking into account the relations (\ref{26}) and (\ref{27}) for the quasiparticle energies $p_{0i}$  at $\Delta=0$ or $M=0$, we have
\begin{eqnarray}
F_1^T(M)
=F_1(M)&-&T\sum_{\kappa=\pm}\int_{0}^{\infty}\frac{dq}{\pi}\ln\Big (1+e^{-\beta\big |\mu+\nu+\kappa\sqrt{M^2+q^2}\big |}\Big )\nonumber\\&-&T\sum_{\kappa=\pm}\int_{0}^{\infty}\frac{dq}{\pi}\ln\Big (1+e^{-\beta\big |\mu-\nu+\kappa\sqrt{M^2+q^2}\big |}\Big ),\label{027}
\end{eqnarray}
where $F_1(M)$ is given in Eq. (\ref{33}). 
Then, the projection $F_2^T(\Delta)$ of the TDP $\Omega_T^{ren}(M=0,\Delta)$ on the $\Delta$ axis reads
\begin{eqnarray}
F_2^T(\Delta)=F_1^T(\Delta)\Big |_{\nu\longleftrightarrow\nu_5}=F_2(\Delta)&-&T\sum_{\kappa=\pm}\int_{0}^{\infty}\frac{dq}{\pi}\ln\Big (1+e^{-\beta\big |\mu+\nu_5+\kappa\sqrt{\Delta^2+q^2}\big |}\Big )\nonumber\\&-&T\sum_{\kappa=\pm}\int_{0}^{\infty}\frac{dq}{\pi}\ln\Big (1+e^{-\beta\big |\mu-\nu_5+\kappa\sqrt{\Delta^2+q^2}\big |}\Big ), \label{034}
\end{eqnarray}
where $F_2(\Delta)$ is given in Eq. (\ref{34}). By comparing the smallest values of the functions $F_1^T(M)$ and 
$F_2^T(\Delta)$, one can construct the simplest temperature phase diagrams of the model. 

First, let us notice that the duality is exact in the chiral limit at any temperature in the framework of NJL$_2$ model.

The phase diagrams of Fig. 4 are the extensions of some phase portraits at $T=0$ shown in Figs. 1 and 2 to nonzero values of temperature. 
For example, Fig. 4 (left panel) is the temperature extension of a section $\nu=m$
of the left panel diagram of Fig. 1 and shows at what temperatures the phase structure remains nontrivial. The middle diagram of Fig. 4
is the temperature extension of a section $\nu/m=0.9$ of the middle panel at Fig. 1. Finally, in the right panel of Fig. 4 one can see
the temperature extension of a section $\nu_5/m=0.6$ of the right panel of Fig. 2. The dotted lines denote second order phase 
transitions, but one can note that there are regions of first order phase transitions at not very large temperatures denoted by the 
solid lines. The points, where two types of phase transitions meet, are the so-called tricritical points (at the physical point, i.e. 
at nonzero bare quark mass, they correspond to the critical endpoints of a phase diagram).   Recall that at zero temperature all phase transitions were of first order.
So, in Fig. 4 we show three $(\mu,T)$-phase diagrams of the model. Each of them corresponds to a rather high fixed value of the 
chemical potential $\nu\gtrapprox m$, but to substantially not very large fixed values of $\nu_5=0,~0.6m$ and $0.75m$, respectively. 
It is clear that temperature phase diagrams of Fig. 4 support also the fact that chiral isospin chemical potential $\mu_{I5}$
promotes the creation of a phase with condensation of charged pions in dense quark matter (it is the PC$_d$ phase in Fig. 4). 
In addition, it is clear from Fig. 4 that the PC$_d$ phase is quite stable with respect to
temperature effects and can be realized in dense quark matter up to rather high temperatures. In Fig. 4 (middle and right panels) 
one can see that PC$_d$ phase persists up to $T=0.2m-0.35m$ respectively, which are comparatively high temperatures 
(if one consider $m$ to be constituent quark mass value in vacuum around 300 MeV, then PC$_d$ phase corresponds to temperatures as 
high as 100 MeV). This part of phase diagram is rather ineresting because these conditions are realized in various physical scenarios such as in just born neutron stars (proto-neutron stars) \cite{Pons:1998mm}, supernovae \cite{Fischer:2011zj} 
and neutron star mergers \cite{Bauswein:2012ya} as well as in heavy ion collisions  \cite{Dexheimer:2017ecc}.

Let us recall that, according to the previous study of quark matter in the framework of the NJL$_2$ model (1) (see in Ref. \cite{ekk}) 
at sufficiently high baryon densities, the PC$_{d}$ and CSB$_{d}$ phases could be realized there only in narrow regions of the 
$(\mu_I,\mu_{I5})$-phase diagram, namely (i) at rather large $\mu_{I5}$ and small $\mu_{I}$ and (ii) at rather large $\mu_{I}$ and 
rather small $\mu_{I5}$, respectively. Other regions of the phase diagram were occupied by either the phases without baryon density or
symmetric phase. According to this analysis, in cores of neutron stars, where there is large isospin imbalance $\mu_{I}$, 
only the CSB$_{d}$ phase can be realized at not very high chiral imbalance  $\mu_{I5}$, and PC$_{d}$ phase was quite unlikely to be 
present (because it required rather low isospin imbalance, which is not the case for the neutron stars). 
In a similar way, the NJL$_2$ analysis forbade the realization of the PC$_d$ phase
in heavy ion collisions where, if one assumes that chiral imbalance is generated due to chiral
separation effect and is rather large, one needs large baryon chemical potential $\mu_{B}$ that can be attained only at comparatively
low energy and in that regime isospin imbalance $\mu_{I}$ can be not small enough for the PC$_d$ phase generation. 
Or if it is more energetic collisions and due to smaller $\mu_B$ probably (let us put aside the magnitude of magnetic field) 
chiral imbalance is small, then isospin imbalance could be not large enough (due to larger energy). 
(Or conversely for CSB$_{d}$ phase, which is probably less interesting but still, if isospin imbalance is large,
then chiral imbalance should be necessarily small.) Of course, these conditions can easily be attained but still, it puts some 
restrictions on the generation of PC$_d$ in heavy ion collisions.  In this
paper we show that in the framework of NJL$_2$ at large baryon density (i.e. at large values of $\mu_{B}$) the regions
occupied by PC$_d$ and CSB$_d$ phases are rather large (two bands) and they are not confined to the small areas near
$\mu_{I5}$ and $\mu_{I}$ axes of the $(\mu_I,\mu_{I5})$-phase diagram. This leads to the possibility of generation of charged pion 
condensation in dense quark matter with large isospin imbalance, for example in cores of neutron stars. The possibility of 
generation of PC$_d$ phase in heavy ion collisions is less bounded by various constraints and the values of different imbalances 
either. So one can see that the possibility of charged pion condensation in dense quark matter was underestimated in Ref. \cite{ekk}
and is actually more considerable result of the NJL$_2$ model predictions.

 Finally, recall that in
the model (1) there is a duality between the CSB and charged PC phenomena. In this case it is possible, relying on the phase
portraits of Fig. 4, to get an idea of the $(\mu,T)$-phase structure of the model, when $\nu$ and $\nu_5$ are fixed in the
qualitatively different way than in Fig. 4 area of the chemical potentials. Namely, when $\nu_5$ is fixed at rather high,
but $\nu$ is fixed at not very high values, respectively. For example, applying to Fig. 4 (left panel) the
duality transformation, at which $\nu\leftrightarrow\nu_5$, CSB$\leftrightarrow$PC and CSBd$\leftrightarrow$PCd, one can
obtain the $(\mu,T)$-phase portrait of the model at $\nu_5=m$ and $\nu=0$. In this new $(\mu,T)$-phase diagram, one can
find in the vicinity of the point $\mu=m$ a charged PC phase with nonzero baryon density, which in itself is already an
interesting fact since in this case there is no isotopic asymmetry ($\nu=0$). In a similar way, one can construct the duality mapping of
other phase portraits in Fig. 4.
\begin{figure}
\includegraphics[width=0.323\textwidth]{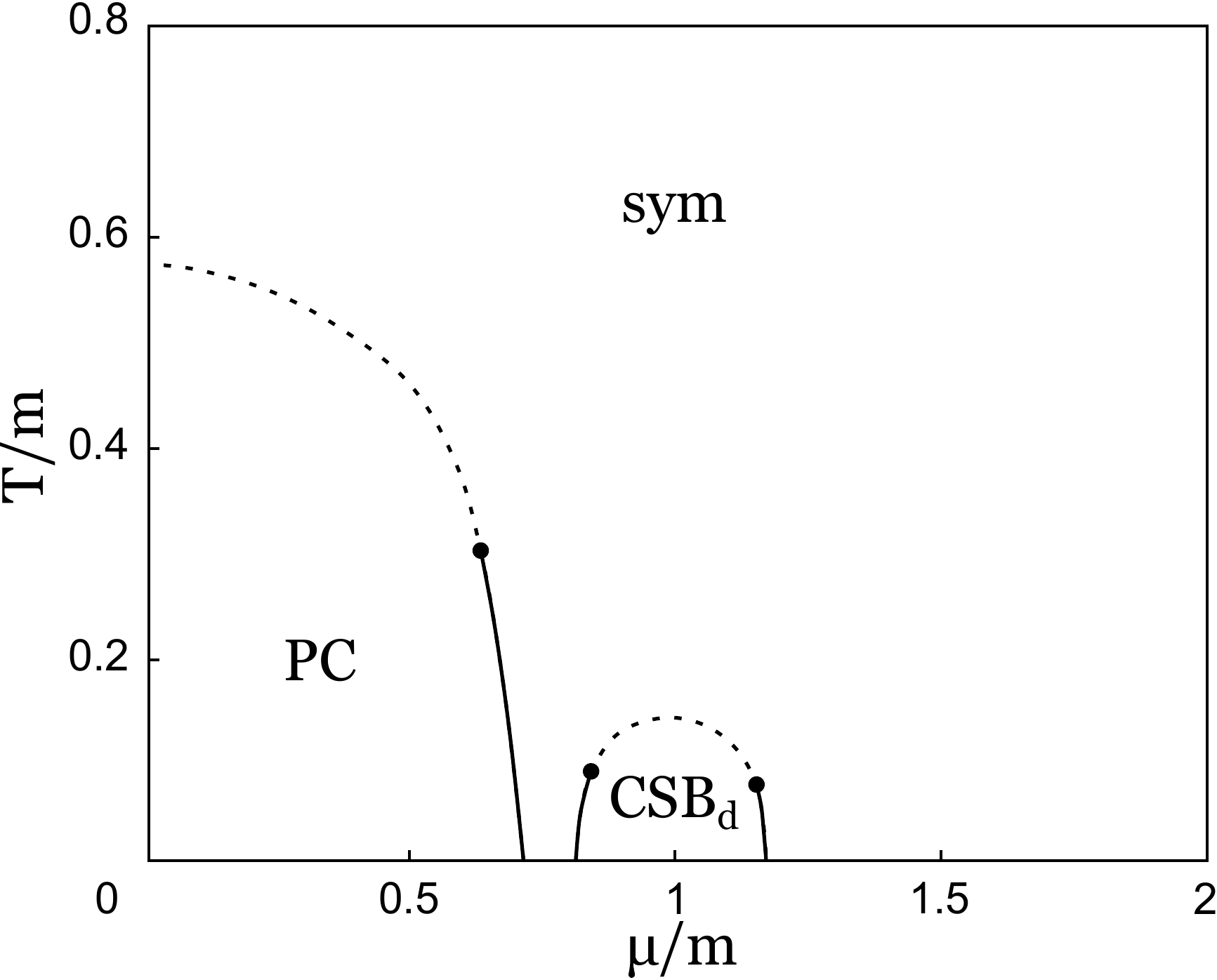}
\includegraphics[width=0.323\textwidth]{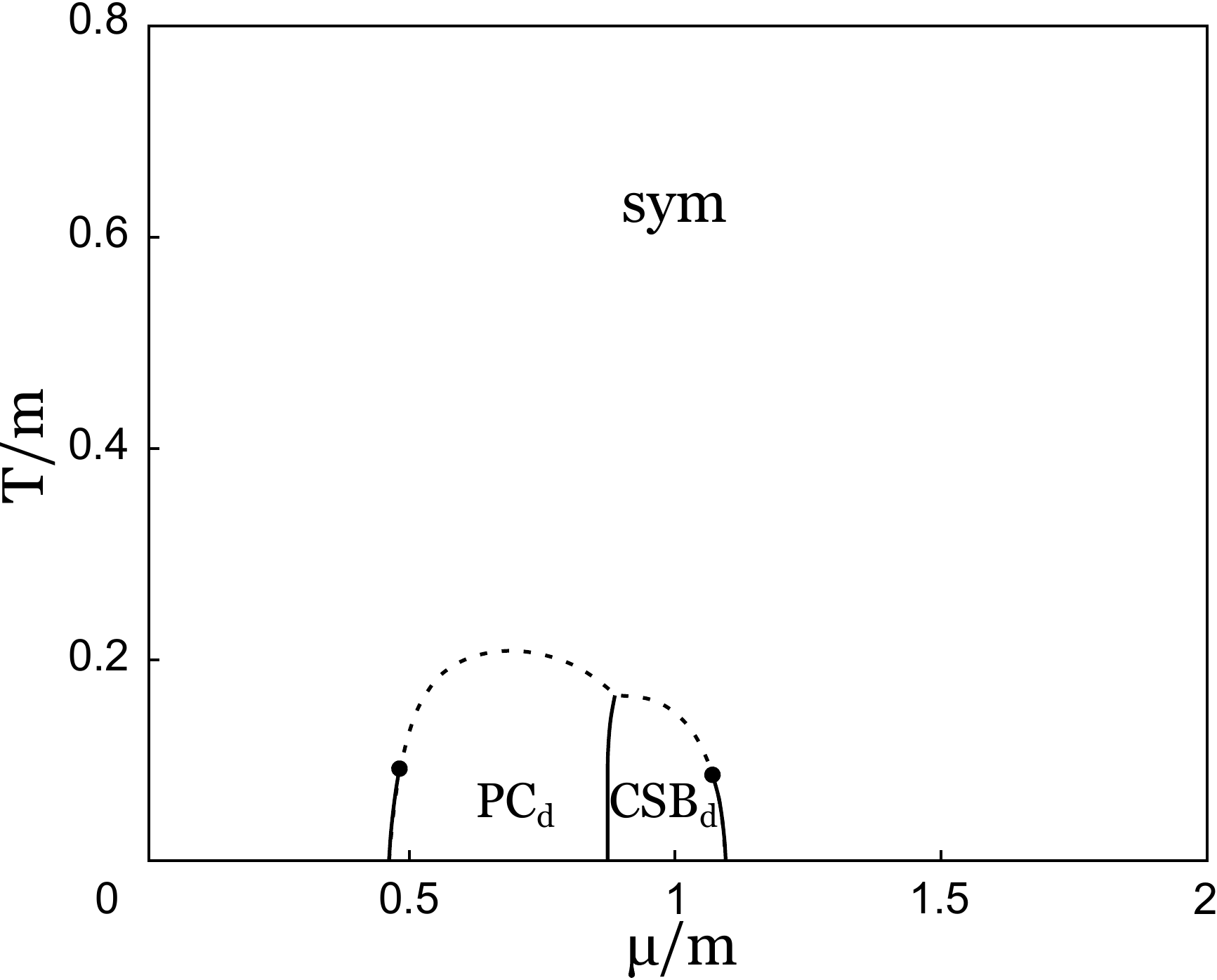}
\includegraphics[width=0.323\textwidth]{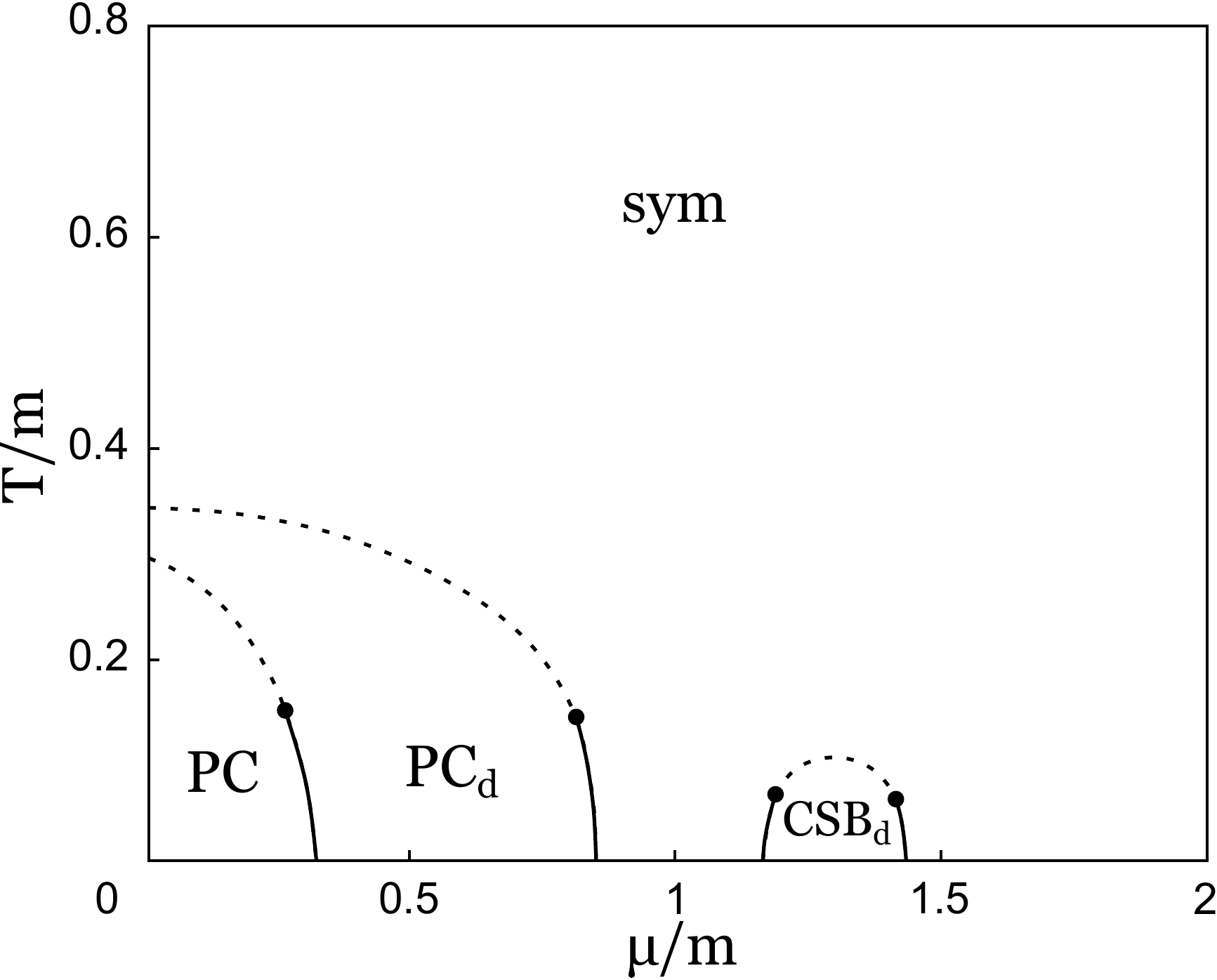}
\caption{The $(T,\mu)$-phase portraits of the model for different
values of the chemical potentials $\nu$ and $\nu_5$: the case
$\nu=m$ and $\nu_5=0$ (left panel), the case  $\nu=0.9m$ and $\nu_5=0.75m$ (middle panel) and the case $\nu=1.3m$ and $\nu_5=0.6m$ 
(right panel). Dashed lines are the second order phase transition curves, solid lines are the first order phase transition curves. 
All the notations are the same as in Fig. 1. The dots are the tricritical points where 
phase transitions of two types meet.}
\end{figure}

\section{Summary and conclusions}

In this paper the phase structure of the massless NJL$_2$ model (1) with two
quark flavors is investigated in the large-$N_c$ limit in the presence
of baryon $\mu_B$, isospin $\mu_I$ and chiral isospin $\mu_{I5}$ chemical potentials.  For the particular case with 
$\mu_{I5}=0$, the task was solved earlier in Refs \cite{ekkz,massive,ek2}, where it was shown that the toy model (1) 
does not predict a charged PC phase of dense and isotopically asymmetric quark matter. 
The  case $\mu_{I5}\ne 0$ has been actually investigated first of all in Ref. \cite{ekk}, where the phase structure of this
NJL$_2$ model has been studied at $T=0$. It was shown there that the charged
PC phase with nonzero baryon density (this phase is denoted by the symbol PC$_d$), prohibited at $\mu_{I5}=0$, appears at 
rather large values of $\mu_{I5}\ne 0$, hence, it was concluded that  chiral isospin asymmetry (i.e. $\mu_{I5}\ne 0$) can 
generate a charged PC phenomenon in dense quark matter. (Note that two other known 
factors promoting a charged PC phase in dense quark matter are finite volume \cite{ekkz} and possible spatial 
inhomogeneity of condensates \cite{gkkz}.)

After this prediction has been made in the framework of NJL$_2$ model (1), the phase structure of a more realistic 
(3+1)-dimensional NJL model with nonzero $\mu_B$, $\mu_I$ and $\mu_{I5}$ was investigated in Refs. \cite{kkz,kkz2} and 
the similar results were obtained, i.e., it was shown in these papers that chiral isotopic chemical potential $\mu_{I5}$ 
(as well as the chiral chemical potential $\mu_5$ \cite{kkz})
generates charged PC phenomenon in dense quark matter. The results were qualitatively the similar, 
but one feature was quite different, namely it was shown that PC$_d$ phase can be generated only at rather small isospin 
chemical potential $\mu_{I}$ values in the NJL$_2$ model and in a wide range of $\mu_{I}$ values in the NJL$_4$ model. 

In the present paper we note that in Ref. \cite{ekk} a miscalculation was performed in the calculation of the expression for 
the TDP of the NJL$_2$ model, and here the corrected and enlarged consideration is presented. In addition, we included 
in the consideration the case of  nonzero temperature. It is shown that although 
the right expression for the TDP is very different from the original one of the paper \cite{ekk} (the right one is much 
simpler), qualitatively the basic features and the main predictions of the NJL$_2$ model analysis turned out to be similar. 

$\bullet$ Namely, the main conclusion that chiral isospin imbalance generates charged pion condensation in dense and cold ($T=0$)
quark matter remains valid. Moreover, it is quite interesting that 
the PC$_d$ phase can occur in dense, {\it isotopically symmetric} baryonic medium, $\nu=0$, but at the same time 
$\nu_5$ should be non-zero (see the left panel of Fig. 2).

$\bullet$ This generation of the PC$_d$ phase by $\nu_5$ does not just remain valid, but 
it is even greatly enhanced and the PC$_d$ phase occupies a long strip along the $\mu_I$ axis of different phase 
portraits (see in Figs. 1-3) of the NJL$_2$ model, instead of a small and compact part of the phase diagram at not 
too large values of isospin chemical potential in Ref. \cite{ekk}. 

$\bullet$ The possibility for the generation of charged pion
condensation in dense quark matter was underestimated in Ref. \cite{ekk} and is actually more considerable. 
It can be realized 
in real physical situations such as heavy ion collisions or compact stars (see the discussion in the section V).

$\bullet$ Moreover, it is also shown in the present paper that in the leading order of the large-$N_c$ expansion there is a 
duality correspondence between CSB and charged PC phenomena. The simplest and most obvious manifestation of this 
phenomenon is that in each $(\mu_I,\mu_{I5})$-phase diagram of the model (see, e.g., in Fig. 3), each CSB phase is located 
mirror-symmetrically to some charged PC phase with respect to the line $\mu_I=\mu_{I5}$ (and vice versa). 

$\bullet$ It has been shown that in the chiral limit duality stays the exact symmetry of the phase diagram of dense quark 
matter even at very high temperatures as long as the low-energy effective NJL model description is valid,
and can be used for phase structure investigations.



$\bullet$ It has been also demonstrated that the predicted generation of charged pion condensation by chiral isospin imbalance in 
dense and cold ($T=0$)
quark matter remains valid even at the temperatures as high as several dozens or even a hundred MeV. 
These conditions are rather pertinent in proto-neutron stars \cite{Pons:1998mm}, supernovae \cite{Fischer:2011zj} 
and neutron star mergers \cite{Bauswein:2012ya} as well as in heavy ion collisions  \cite{Dexheimer:2017ecc}. 

So this consideration shows that the possibility of charged pion condensation in dense quark matter at zero temperature was previously underestimated and it is more likely to happen in compact stars. Additionally, it was shown that this phenomenon can take place even at rather high temperatures, which makes it 
feasible in more various physical situations such as heavy ion collisions, just born neutron stars (proto-neutron stars), supernovas as well as neutron star mergers. 

Taking into account earlier investigations \cite{kkz}, we see that within the framework of two different effective models, 
both in NJL$_4$ and NJL$_2$ models, quite the similar properties (i.e., the similar phase portraits)
of dense and chirally asymmetric quark matter are predicted. This fact increases our confidence that in the cores of 
neutron stars, where $\mu_{B}$ and $\mu_{I}$ can reach significant values, 
the effect of charged PC can be realized. 
Then, to explain the similarity of the phase diagrams of these two effective models, two circumstances can be taken into 
account. (i) Under the influence of a strong magnetic field in the depths of neutron stars, regions with a nonzero chiral 
isotopic density (with $\mu_{I5}\ne 0$) can appear (for more details, see Appendix A of Refs. \cite{kkz2,kkzjhep}). (ii) Just 
in the presence of a strong magnetic field, the dynamics of fermion pairing is essentially (1 + 1)-dimensional \cite{Gusynin:1994xp}. 
Therefore, the phase portraits of dense quark matter at $\mu_{I5}\ne 0$ obtained in framework of the NJL$_4$ and NJL$_2$ 
models are qualitatively the same. However, it is worth noting that (1 + 1)-dimensional models are not comprehensive. 
Thus, phase transformations in real dense media 
with a strong magnetic field are still available for description in a two-dimensional framework, which cannot be said 
of many other physical properties. Therefore, the influence of the charged PC phase on various processes in neutron stars,
such as transport processes etc., cannot be considered within 
NJL$_2$ model. It is not the subject of the present paper, and will be investigated in the framework of a more realistic 
model elsewhere.

\section{Acknowledgments}
R.N.Z. is grateful for support of the Foundation for the Advancement of Theoretical Physics and Mathematics BASIS grant
and Russian Science Foundation under the grant  N\textsuperscript{\underline{\scriptsize o}} 19-72-00077

\appendix
\section{Evaluation of the roots of the polynomial $P_4(p_0)$ (\ref{91}) }
\label{ApB}


It is very convenient to present the fourth-order polynomial (\ref{91}) of the variable $\eta\equiv p_0+\mu$  as a product of two second-order polynomials (this way is proposed in \cite{Birkhoff}), i.e. we assume that
\begin{eqnarray}
\eta^4&-&2a \eta^2-b \eta +c = (\eta^2 + r\eta + q)(\eta^2 - r\eta + s)\label{B1}\\
&&=\left [\left (\eta+\frac r2\right )^2+q-\frac{r^2}{4}\right ]\left [\left
(\eta-\frac r2\right )^2+s-\frac{r^2}{4}\right ]\equiv
(\eta-\eta_{1})(\eta-\eta_{2})(\eta-\eta_{3})(\eta-\eta_{4}),\label{B3}
\end{eqnarray}
where $r$, $q$ and $s$ are some real valued quantities, such that
(see the relations (\ref{10})):
\begin{eqnarray}
-2a& \equiv&-2(M^2+\Delta^2+p_1^2+\nu^2+\nu_{5}^2)= s+q-r^2;~~ -b\equiv -8p_1\nu\nu_{5}= rs-qr;\nonumber\\
c&\equiv& a^2-4p_1^2(\nu^2+\nu_5^2)-4M^2\nu^2-4\Delta^2\nu_5^2-4\nu^2\nu_5^2=sq.
\label{B4}
\end{eqnarray}
In the most general case, i.e. at $M\ge 0$, $\Delta\ge 0$, $\nu\ge 0$,$\nu_5\ge 0$ and arbitrary values of $p_1$, one can solve the system of equations (\ref{B4}) with respect to $q,s,r$ and find
\begin{eqnarray}
q=\frac 12 \left (-2a +R+\frac{b}{\sqrt{R}}\right ),~~ s=\frac 12 \left
(-2a +R-\frac{b}{\sqrt{R}}\right ),~~r=\sqrt{R},\label{B5}
\end{eqnarray}
where $R$ is an arbitrary real positive solution of the equation
\begin{eqnarray}
\label{B6} X^3 + AX=BX^2 + C
\end{eqnarray}
with respect to a variable $X$, and
\begin{eqnarray}
A&=& 4a^2-4c=16\Big[\nu_5^2\Delta^2+M^2\nu^2+
\nu_5^2\nu^2+p_1^2(\nu^2+\nu_5^2)\Big],\nonumber\\
B&=&4a =4(M^2+\Delta^2+\nu^2+\nu_5^2+p_1^2),~~ C=b^2=(8\nu_5\nu
p_1)^2. \label{B7}
\end{eqnarray}
Finding (numerically) the quantities $q$, $s$ and $r$, it is possible to obtain from (\ref{B3}) the roots $\eta_i$:
\begin{eqnarray}
\eta_{1}=-\frac{r}{2}+\sqrt{\frac{r^2}{4}-q},~~\eta_{2}=\frac{r}{2}+
\sqrt{\frac{r^2}{4}-s},~~\eta_{3}=-\frac{r}{2}-\sqrt{\frac{r^2}{4}-q},~~\eta_{4}=\frac{r}{2}-\sqrt{\frac{r^2}{4}-s}.\label{B41}
\end{eqnarray}
Numerical investigation shows that in the most general case the discriminant of the third-order algebraic equation (\ref{B6}), i.e. the quantity $18ABC-4B^3C+A^2B^2-4A^3-27C^2$, is always nonnegative. So the equation (\ref{B6}) vs $X$ has three real solutions $R_1,R_2$ and $R_3$ (this fact is presented in \cite{Birkhoff}). Moreover, since the coefficients $A$, $B$ and $C$ (\ref{B7})
are nonnegative, it is clear that, due to the form of equation (\ref{B6}), all its roots $R_1$, $R_2$ and $R_3$ are also nonnegative quantities (usually, they are positive and different). So we are free to choose the quantity $R$ from (\ref{B5}) as one of the positive solutions $R_1$, $R_2$ or $R_3$. In each case, i.e. for
$R=R_1$, $R=R_2$, or $R=R_3$, we will obtain the same set of 
roots (\ref{B41}) (possibly rearranged), which depends only on $\nu$, $\nu_5$, $M$, $\Delta$ and $p_1$, and does not depend on 
the choice of $R$. Due to the relations (\ref{B3})-(\ref{B41}), one can find numerically (at fixed values of  
$\mu$, $\nu$, $\nu_5$, $M$, $\Delta$ and $p_1$) the roots $\eta_i=p_{0i}+\mu$ (\ref{B41}) and, as a result, investigate 
numerically the TDP (\ref{28}). 

Analyzing the relations (\ref{B1})-(\ref{B41}), we can draw another important conclusion concerning the properties of the roots 
$\eta_i$ of the polynomial (\ref{B1}) with respect to the duality transformations ${\cal D}$ (\ref{16}). It is clear that the 
coefficients of this polynomial are invariant under the action of the ${\cal D}$, so each root $\eta_i$ (\ref{B41}) is transformed 
under the dual mapping ${\cal D}$ to itself, if the quantity $R$ from (\ref{B5}) remains intact under the action of ${\cal D}$ 
(see below in Eq. (\ref{B26})). Note that in our work the quantity $R$ is chosen with exactly this property.
However, in the most general case (when $R$ is selected to be not invariant under the ${\cal D}$ transformation) the set of 
roots $\eta_i$ also remains unchanged (with the cost of a possible rearrangement 
of these roots) under the action of ${\cal D}$. Hence, the TDPs of Eqs. (29), (35) etc. are invariant with respect to the duality 
transformation (\ref{16}). 
 
On the basis of the relations (\ref{B1})-(\ref{B41}) let us consider the asymptotic behavior of the quasiparticle energies 
$p_{0i}$ at $p_1\to\pm\infty$. First of all, we start from the asymptotic analysis of the roots $R_{1,2,3}$ of the equation 
(\ref{B6}) at $p_1\to\pm\infty$,
\begin{eqnarray}
\label{B701}
R_1&=&4\nu^2-\frac{4\Delta^2\nu^2}{p_1^2}
+{\cal O}\big (1/p_1^4\big ),\\
\label{B7001}
R_2&=&4\nu_5^2-\frac{4M^2\nu_5^2}{p_1^2}
+{\cal O}\big (1/p_1^4\big ),\\
R_3&=&4p_1^2+4(M^2+\Delta^2)+\frac{4(\nu_5^2M^2+\nu^2\Delta^2)}{p_1^2}
+{\cal O}\big (1/p_1^4\big ). \label{B71}
\end{eqnarray}
It is clear from these relations that $R_3$ is invariant under the duality transformation (\ref{16}), whereas $R_1\leftrightarrow R_2$.
Then, using for example $R_3$ (\ref{B71}) as the quantity $R$ in Eqs. (\ref{B5}) and (\ref{B41}), one can get the asymptotics of 
the quasiparticle energies $p_{0i}\equiv \eta_i-\mu$ at $p_1\to\pm\infty$,
\begin{eqnarray}
p_{01}&=&-|p_1|-\mu+|\nu_5-\nu|-\frac{\Delta^2+M^2}{2|p_1|} +{\cal O}\big
(1/p_1^2\big ),~~ p_{02}=|p_1|-\mu+\nu_5+\nu+\frac{\Delta^2+M^2}{2|p_1|}
+{\cal O}\big (1/p_1^2\big ),\nonumber\\
p_{03}&=&-|p_1|-\mu-|\nu_5-\nu|-\frac{\Delta^2+M^2}{2|p_1|} +{\cal O}\big
(1/p_1^2\big ),~~ p_{04}=|p_1|-\mu-\nu_5-\nu+\frac{\Delta^2+M^2}{2|p_1|}
+{\cal O}\big (1/p_1^2\big ).\label{B26}
\end{eqnarray}
Finally, it follows from (\ref{B26}) that at $p_1\to\pm\infty$
\begin{eqnarray}
|p_{01}|+|p_{02}|+|p_{03}|+|p_{04}|=4|p_1|+\frac{2(\Delta^2+M^2)}{|p_1|}
+{\cal O}\big (1/p_1^2\big ).\label{B9}
\end{eqnarray}
For the purposes of the renormalization of the TDP (\ref{28}), it is
very important that the leading terms of this asymptotic behavior do
not depend on different chemical potentials, i.e. the quantity
$\sum_{i=1}^4|p_{0i}|$ at $\mu=\nu=\nu_5=0$ has the same asymptotic
expansion (\ref{B9}). Furthermore, there is an exact expression for this sum,
\begin{eqnarray}
\big (|p_{01}|+|p_{02}|+|p_{03}|+|p_{04}|\big )\Big |_{\mu=\nu=\nu_5=0}=
4\sqrt{M^2+\Delta^2+p_1^2}.\label{B10}
\end{eqnarray}
We would like to emphasize once again that the
asymptotic behavior (\ref{B9}) does not depend on which of the roots
$R_1$, $R_2$ or $R_3$ of the equation (\ref{B6}) is taken as the
quantity $R$ in the relations (\ref{B5}). Finally, it is important to note that the quantity $\sum_{i=1}^4|p_{0i}|$ is not an 
even function with respect to the momentum $p_1$. (This statement is supported, in particular, by the relation (\ref{36}) of the next 
Appendix B.) Indeed, the polynomial (\ref{B1}) is not invariant under the transformation $p_1\to -p_1$ (since its coefficient $b$
changes). So both each root $\eta_i$ and the sum $\sum_{i=1}^4|p_{0i}|$ are changed under this transformation. (In contrast, if $\nu=0$ or 
$\nu_5=0$ then this sum is an even function vs $p_1$.)
Unfortunately, the opposite, incorrect assertion was made in Ref. \cite{ekk}, which led to an incorrect expression for the TDPs $F_1(M)$ and $F_2(\Delta)$. 

\section{Derivation of the relation (\ref{33})}
\label{ApD}
If $\Delta=0$ and $M\ne 0$, then the quasiparticle energies $p_{0i}$ are presented in the expression (\ref{26}). So
\begin{eqnarray}
\big (|p_{01}|+|p_{02}|+|p_{03}|+|p_{04}|\big )\big |_{\Delta=0}&=&\nonumber\\
\sum_{\kappa=\pm}\Big (\left
|-\mu+\kappa\nu+\sqrt{M^2+(p_1+\kappa\nu_5)^2}\right |&&\hspace{-4mm}+\left
|-\mu+\kappa\nu-\sqrt{M^2+(p_1+\kappa\nu_5)^2}\right |\Big )\nonumber\\
=2\sum_{\kappa=\pm}\Bigg\{\sqrt{M^2+(p_1+\kappa\nu_5)^2}&&\hspace{-4mm}+
\Big (\mu-\kappa\nu-\sqrt{M^2+(p_1+\kappa\nu_5)^2}\Big )
\theta \Big (\mu-\kappa\nu-\sqrt{M^2+(p_1+\kappa\nu_5)^2}\Big )\nonumber\\
+\Big (\kappa\nu-\mu&&\hspace{-4mm}-\sqrt{M^2+(p_1+\kappa\nu_5)^2}\Big )
\theta \Big (\kappa\nu-\mu-\sqrt{M^2+(p_1+\kappa\nu_5)^2}\Big )\Bigg\}, \label{36}
\end{eqnarray}
where we have took into account the well-known relations $|x|=x\theta
(x)-x\theta (-x)$ and $\theta (x)=1-\theta(-x)$. Hence, the expression (\ref{35}) at $\Delta=0$ and $M\ne 0$ can be presented in the following form:
\begin{eqnarray}
F_1(M)\equiv\Omega^{ren} (M,\Delta=0)&=&-\frac{M^2}{2\pi}+\frac{M^2}{2\pi}\ln\left
(\frac{M^2}{m^2}\right )-U-V, \label{D1}
\end{eqnarray}
where
\begin{eqnarray}
U&=&\int_{-\infty}^\infty\frac{dp_1}{2\pi}\Big\{\sqrt{M^2+(p_1+\nu_5)^2}+\sqrt{M^2+(p_1-\nu_5)^2}-2\sqrt{M^2+p_1^2}\Big\}=\frac{\nu_5^2}{\pi},\label{D2}\\
V&=&\sum_{\kappa=\pm}\int_{-\infty}^\infty\frac{dp_1}{2\pi}\Big\{\Big (\mu-\kappa\nu-\sqrt{M^2+(p_1+\kappa\nu_5)^2}\Big
) \theta\Big (\mu-\kappa\nu-\sqrt{M^2+(p_1+\kappa\nu_5)^2}\Big )\nonumber\\&&~~~~~~~+\Big
(\kappa\nu-\mu-\sqrt{M^2+(p_1+\kappa\nu_5)^2}\Big ) \theta\Big
(\kappa\nu-\mu-\sqrt{M^2+(p_1+\kappa\nu_5)^2}\Big )\Big\}\label{D20}\\
&=&\int_{-\infty}^\infty\frac{dp_1}{2\pi}\Big (\mu-\nu-\sqrt{M^2+(p_1+\nu_5)^2}\Big ) \theta\Big (\mu-\nu-\sqrt{M^2+(p_1+\nu_5)^2}\Big )\nonumber\\&+&\int_{-\infty}^\infty\frac{dp_1}{2\pi}\Big
(\nu-\mu-\sqrt{M^2+(p_1+\nu_5)^2}\Big ) \theta\Big
(\nu-\mu-\sqrt{M^2+(p_1+\nu_5)^2}\Big )\nonumber\\
&+&\int_{-\infty}^\infty\frac{dp_1}{2\pi}\Big (\mu+\nu-\sqrt{M^2+(p_1-\nu_5)^2}\Big
) \theta\Big (\mu+\nu-\sqrt{M^2+(p_1-\nu_5)^2}\Big ).\label{D3}
\end{eqnarray}
Notice that a calculation of the convergent improper integral $U$
(\ref{D2}) can be found, e.g., in Appendix C of \cite{ekkz}. Moreover,
when summing in (\ref{D20}) over $\kappa =\pm$, we took into account
that $\mu\ge 0$ and $\nu\ge 0$. So there are only three integrals in
the expression (\ref{D3}). Due to the presence of the step function
$\theta (x)$, each integral in (\ref{D3}) is indeed a proper one. Let
us denote the sum of the first two integrals of (\ref{D3}) as $V_1$
and the last integral as $V_2$, i.e. $V=V_1+V_2$. Then, it is evident that
\begin{eqnarray}
V_1&=&\int_{-\infty}^\infty\frac{dp_1}{2\pi}\Big (|\mu-\nu|-\sqrt{M^2+(p_1+\nu_5)^2}\Big ) \theta\Big (|\mu-\nu|-\sqrt{M^2+(p_1+\nu_5)^2}\Big ),\label{D4}\\
V_2&=&\int_{-\infty}^\infty\frac{dp_1}{2\pi}\Big (\mu+\nu-\sqrt{M^2+(p_1-\nu_5)^2}\Big ) \theta\Big (\mu+\nu-\sqrt{M^2+(p_1-\nu_5)^2}\Big ).\label{D5}
\end{eqnarray}
Now let us turn in these relations to the integration over the region $p_1\in (0,+\infty)$. Then
\begin{eqnarray}
V_1&=&\int_{0}^\infty\frac{dp_1}{2\pi}\Big (|\nu-\mu|
-\sqrt{M^2+(p_1+\nu_5)^2}\Big )
\theta \Big (|\nu-\mu|-\sqrt{M^2+(p_1+\nu_5)^2}\Big )\nonumber\\
&+&\int_{0}^\infty\frac{dp_1}{2\pi}\Big (|\nu-\mu|
-\sqrt{M^2+(p_1-\nu_5)^2}\Big )
\theta \Big (|\nu-\mu|-\sqrt{M^2+(p_1-\nu_5)^2}\Big )\Bigg\},\label{D6}\\
V_2&=&\int_{0}^\infty\frac{dp_1}{2\pi}\Big (\mu+\nu-\sqrt{M^2+(p_1-\nu_5)^2}\Big )
\theta \Big (\mu+\nu-\sqrt{M^2+(p_1-\nu_5)^2}\Big )\nonumber\\
&+&\int_{0}^\infty\frac{dp_1}{2\pi}\Big (\mu+\nu-\sqrt{M^2+(p_1+\nu_5)^2}\Big )
\theta \Big (\mu+\nu-\sqrt{M^2+(p_1+\nu_5)^2}\Big )\equiv V_{2-}+V_{2+},
 \label{D7}~~~~~~~~~~~D7 
\end{eqnarray}
where $V_{2-}$ and $V_{2+}$ denote the first and the second integral of Eq. (\ref{D7}), respectively. Carring out in the integrals of Eq. (\ref{D7}) the change of variables, $q=p_1+\nu_5$ for $V_{2+}$ and  $q=p_1-\nu_5$ for $V_{2-}$, respectively, we have
\begin{eqnarray}
V_{2+}&=&\int_{\nu_5}^\infty\frac{dq}{2\pi}\Big (\mu+\nu-\sqrt{M^2+q^2}\Big ) \theta\Big (\mu+\nu-\sqrt{M^2+q^2}\Big )\nonumber\\
&=&\left (\int_{0}^\infty-\int^{\nu_5}_0\right )\frac{dq}{2\pi}\Big (\mu+\nu-\sqrt{M^2+q^2}\Big ) \theta\Big (\mu+\nu-\sqrt{M^2+q^2}\Big ),\label{D8}~~~~~~D8\\
V_{2-}&=&\int_{-\nu_5}^\infty\frac{dq}{2\pi}\Big (\mu+\nu-\sqrt{M^2+q^2}\Big ) \theta\Big (\mu+\nu-\sqrt{M^2+q^2}\Big )\nonumber\\
&=&\left (\int_{0}^\infty+\int^{\nu_5}_0\right )\frac{dq}{2\pi}\Big (\mu+\nu-\sqrt{M^2+q^2}\Big ) \theta\Big (\mu+\nu-\sqrt{M^2+q^2}\Big ).\label{D9}~~~~~~~~~D9
\end{eqnarray}
Hence, 
\begin{eqnarray}
&&V_2\equiv V_{2+}+V_{2-}=\int_{0}^\infty\frac{dq}{\pi}\Big (\mu+\nu-\sqrt{M^2+q^2}\Big ) \theta\Big (\mu+\nu-\sqrt{M^2+q^2}\Big )\nonumber\\&=&\frac{\theta (\mu+\nu-M)}{2\pi}\left ((\mu+\nu)\sqrt{(\mu+\nu)^2-M^2}-M^2\ln\frac{\mu+\nu+\sqrt{(\mu+\nu)^2-M^2}}{M}\right ).\label{D10}~~~~~~D10
\end{eqnarray}
In a similar way it is possible to show that
\begin{eqnarray}
V_1&=&\int_{0}^\infty\frac{dq}{\pi}\Big (|\mu-\nu|-\sqrt{M^2+q^2}\Big ) \theta\Big (|\mu-\nu|-\sqrt{M^2+q^2}\Big )\nonumber\\
&=&\frac{\theta (|\mu-\nu|-M)}{2\pi}\left (|\mu-\nu |\sqrt{(\mu-\nu)^2-M^2}-M^2\ln\frac{|\mu-\nu|+\sqrt{(\mu-\nu)^2-M^2}}{M}\right ).\label{D11}~~~~~~D11
\end{eqnarray} 
Since $V=V_1+V_2$ and $U$ is given in Eq. (\ref{D2}), we have for the TDP $F_1(M)$  (\ref{D1}) the expression (\ref{33}).

\end{document}